\documentclass[10pt,letterpaper,final,twocolumn,journal]{IEEEtran}
\pdfoutput=1
\usepackage{setspace}
\usepackage{nomencl}
\makenomenclature
\usepackage{graphicx}
\usepackage{algorithm}
\usepackage{algorithmic}
\usepackage{epstopdf}
\usepackage{epsfig}
\usepackage[cmex10]{amsmath}
\usepackage{cite}
\usepackage{amssymb}
\usepackage[usenames,dvipsnames]{color} 
\usepackage{multirow}
\usepackage{array}
\long\def\symbolfootnote[#1]#2{\begingroup\def\thefootnote{
\fnsymbol{footnote}}\footnote[#1]{#2}\endgroup}
\usepackage[table]{xcolor}
\usepackage{tabularx}
\usepackage{chngcntr}
\usepackage{booktabs}
\usepackage{caption}
\usepackage{subcaption}
\usepackage{float}
\usepackage{url}

\begin{document}

\title{{Near-Hashing-Bound Multiple-Rate Quantum Turbo Short-Block Codes}}
\author{Daryus~Chandra,
Zunaira~Babar,
Soon~Xin~Ng,
and~Lajos~Hanzo
\thanks{The authors are with the School of Electronics and Computer Science, University of Southampton, Southampton, SO17 1BJ, UK (email: \{dc2n14,~zb2g10,~sxn,~lh\}@ecs.soton.ac.uk).}
\thanks{The financial support of the EPSRC under the grant EP/L018659/1 and the COALESCE project, that of the European Research Council, Advanced Fellow Grant QuantCom and that of the Royal Society's Global Research Challenges Fund (GRCF) is gratefully acknowledged. Additionally, the authors acknowledge the use of the IRIDIS High Performance Computing Facility, and associated support services at the University of Southampton, in the completion of this work. 
} 
}

\maketitle

\begin{abstract} 
Quantum stabilizer codes (QSCs) suffer from a low quantum coding rate, since they have to recover the quantum bits (qubits) in the face of both bit-flip and phase-flip errors. In this treatise, we conceive a low-complexity concatenated quantum turbo code (QTC) design exhibiting a high quantum coding rate. The high quantum coding rate is achieved by combining the quantum-domain version of short-block codes (SBCs) also known as single parity check (SPC) codes as the outer codes and quantum unity-rate codes (QURCs) as the inner codes. Despite its design simplicity, the proposed QTC yields a near-hashing-bound error correction performance. For instance, compared to the best half-rate QTC known in the literature, namely the QIrCC-QURC scheme, which operates at the distance of $D = 0.037$ from the quantum hashing bound, our novel QSBC-QURC scheme can operate at the distance of $D = 0.029$. It is worth also mentioning that this is the first instantiation of QTCs capable of adjusting the quantum encoders according to the quantum coding rate required for mitigating the Pauli errors given the different depolarizing probabilities of the quantum channel.
\end{abstract}

\begin{IEEEkeywords}
quantum error detection codes, quantum error correction codes, concatenated codes, iterative decoding, topological codes, quantum turbo codes, short-block codes.
\end{IEEEkeywords}

\begin{table*}[ht!]
\centering
\begin{tabular}{ll}
\hline
& List of Acronyms \\
\hline
CNOT & Controlled-NOT \\
CSS & Calderbank-Shor-Steane \\
EXIT & EXtrinsic Information Transfer\\
PCM & Parity Check Matrix \\ 
QBCH & Quantum Bose-Chaudhuri-Hocquenghem \\
QBER & QuBit Error Ratio \\
QCC & Quantum Convolutional Code\\
QEDC & Quantum Error Detection Code\\
QECC & Quantum Error Correction Code \\
QIrCC & Quantum Irregular Convolutional Code\\
QSC & Quantum Stabilizer Code \\
QSBC & Quantum Short-Block Code\\
QTECC & Quantum Topological Error Correction Code\\
QTC & Quantum Turbo Code\\
QURC & Quantum Unity-Rate Code\\
SBC & Short-Block Code\\
SPC & Single Parity Check\\
\hline
\end{tabular}
\end{table*}

\section{Introduction}
\label{Introduction}

Quantum stabilizer codes (QSCs)~\cite{gottesman1997stabilizer, shor1995scheme, steane1996multiple, steane1996error, calderbank1996good, calderbank1997quantum, calderbank1998quantum} are capable of estimating both the number and the position of quantum bit (qubit) errors without collapsing the quantum state of physical qubits into their classical state. Hence, they can be viewed as the syndrome-based quantum error correction codes (QECCs). However, the QSCs suffer from a lower quantum coding rate $(r_Q)$ than their classical counterparts, since they have to tackle not only the bit-flip $(\mathbf{X})$ errors but also the phase-flip $(\mathbf{Z})$ errors~\cite{chandra2017quantum1}. Motivated by~\cite{abbara2011quantum}, where it has been demonstrated that concatenating a quantum linear block code with a unity-rate quantum convolutional code (QCC), which is also referred to as a quantum unity-rate code (QURC), facilitates the soft-decision-aided iterative decoding. Hence, a dramatic performance improvement may be attained without sacrificing the quantum coding rate. However, the proposal of~\cite{abbara2011quantum} suffered from a relatively high error floor, despite relying on a low overall quantum coding rate of $r_Q = 1/8$, when using a block code having a minimum distance of $d = 3$ as the outer code. Additionally, in order to eliminate the error floor, this specific code construction required a doping mechanism for triggering the convergence of iterative decoding, since the QURC utilized a catastrophic encoder structure. Later in~\cite{babar2016serially}, it was shown that by carefully selecting the inner and the outer codes using an extrinsic information transfer (EXIT)-chart-guided code search, a half-rate quantum turbo code (QTC) can be conceived, which is capable of performing relatively close to the quantum hashing bound. Explicitly, this excellent error correction performance was achieved by a half-rate quantum irregular convolutional code (QIrCC) used as the outer code combined with a QURC as the inner code. This specific code is referred to as the QIrCC-QURC scheme.

As an important result in the classical domain, it was shown that an outer code exhibiting a minimum distance of $d = 2$ is capable of guaranteeing the convergence of iterative decoding to a vanishingly low bit error ratio (BER)~\cite{benedetto1998serial, land2005reliability}, as exemplified by the family of single parity check codes (SPCs) or short-block codes (SBCs)~\cite{hanzo2008short, hanzo2009exit}. Furthermore, by exploiting the classical-to-quantum isomorphism~\cite{chandra2017quantum1, chandra2018quantum2}, we can indeed conceive the quantum version of the classical SBCs, which we referred to as quantum short-block codes (QSBCs). As an additional benefit, the QSBCs can also be viewed as quantum topological error correction codes (QTECCs). More specifically, the QTECC construction exhibits an inherent error detection and error correction capability, when the physical qubits are appropriately arranged on a lattice structure. However, most of the conventional techniques of constructing the QTECCs suffer from a low quantum coding rate as well as from the lack of flexibility, when choosing the number of logical qubits and also the quantum coding rate~\cite{chandra2018quantum2}. By contrast, when a similar approach relying on utilizing a lattice structure is invoked for constructing quantum error detection codes (QEDCs), instead of quantum error correction codes (QECCs), we found that the resultant topological QEDCs are flexible and exhibit high quantum coding rates.

Against this background, by amalgamating QURCs and QSBCs, we conceive the family of high-rate low-complexity serial QTCs, which are capable of operating at various quantum coding rates whilst relying on a flexible numbers of logical qubits. As a further benefit, they are capable of approaching the quantum hashing bound. More explicitly, our main contributions are as follows:
\begin{itemize}
\item \textit{We present the general design of high-rate QSBCs exhibiting a minimum distance of $d = 2$, which constitute the family of QEDCs. Explicitly, the proposed QSBC can have a block length as short as four physical qubits ($n = 4$) and it has a quantum coding rate of $r_Q = \frac{k}{k+1}$. As an additional benefit, the QSBCs have a scalable encoder structure and require only localized stabilizer measurements. In addition, we demonstrate that the associated stabilizer measurement can be localized by arranging the physical qubits on a polygon structure, such as a square, hexagon, octagon, etc. Hence, this is the first instantiation of high-rate, scalable, short-length, and high-rate QEDCs.} 
\item \textit{We amalgamate the QSBCs with the QURCs~\cite{babar2016serially} for the sake of constructing soft-decision-aided QECCs without sacrificing the quantum coding rate, which results in a low-complexity high-rate QTC design. We refer to the resultant construction as a QSBC-QURC scheme. Despite having low complexity, the QSBC-QURC scheme is capable of operating close to the achievable quantum hashing bound. Quantitatively, for instance, our simulation results demonstrate that a half-rate QSBC-QURC operates at the distance of $D = 0.029$ from the quantum hashing bound.}
\item \textit{Finally, we conceive the first instantiation of a multi-rate scheme for serial QTCs relying on the flexible scalability of the QSBC-QURC construction. We determine the specific depolarizing probability values at which it is beneficial to switch the quantum coding rate based on the minimum QBER requirement of $10^{-3}$. Finally, we quantify the achievable goodput of the QSBC-QURC schemes conceived.}
\end{itemize}

The rest of this treatise is organized as follows. In Section~\ref{Quantum Short-Block Codes}, we present the general formulation of QSBCs in terms of their code construction, encoder, and stabilizer measurement. This is followed by Section~\ref{Quantum Turbo Code Design Using QSBCs}, where we propose on serial QTCs by utilizing QSBCs as the outer codes and a QURC as the inner code, which we refer to as the QSBC-QURC scheme. We analyze the convergence behavior of iterative-decoding-aided QSBC-QURC scheme using EXIT charts and evaluate its QBER and goodput in Section~\ref{Results and Analysis}. Finally, we conclude the paper and present some possible directions for future research in Section~\ref{Conclusions and Future Research}.

\section{Quantum Short-Block Codes}
\label{Quantum Short-Block Codes}

In this section, we introduce the quantum version of classical SBCs by describing their general construction, the structure of the quantum encoder, as well as the quantum circuit required for the stabilizer measurements. This section will also characterize both the flexibility and scalability of the QSBC encoders and the associated stabilizer measurements, as the natural evolution from their parity check matrix structure. Furthermore, we present a short tutorial on conducting the classical simulation for QSBCs. In order to avoid ambiguity, throughout the rest of this treatise the notation $\mathcal{C}(n,k,d)$ is used to denote the classical error correction code having codeword length of $n$ bits, information word $k$ bits, and a minimum distance of $d$, while the notation $\mathcal{C}[n,k,d]$ is used for the quantum stabilizer code.

\subsection{Codes Construction}
\label{Codes Construction}

The classical SBCs are systematic linear binary block codes $\mathcal{C}\left( n,k,d \right)$, whose generator matrix $\mathbf{G}$ is defined by:
\begin{equation}
\mathbf{G} =  \left[ \mathbf{I}_{k} | \mathbf{P}_{k,(n-k)} \right] =  \left[ \mathbf{I}_{k} | \mathbf{J}_{k,1} \right],
\end{equation}
where $\mathbf{I}_k$ is a $k$-dimension identity matrix, $\mathbf{P}_{(n-k),k}$ is a $(k \times (n-k))$-element binary matrix and $\mathbf{J}_{m,n}$ is a matrix with all-one $(m \times n)$-element. Hence, the parity check matrix (PCM) $\mathbf{H}$ of a systematic linear block code is encapsulated in
\begin{equation}
\mathbf{H} = \left[ \mathbf{I}_{n-k} | \mathbf{P}^T \right].
\end{equation}
Therefore, the PCM $\mathbf{H}$ of the classical SBCs is given by
\begin{equation}
\mathbf{H} = \mathbf{J}_{1,n},
\end{equation}
which is an all-one $(1 \times n)-$element matrix. Finally, the resultant coding rate is given by 
\begin{equation}
r = \frac{k}{n} = \frac{k}{k+1} = \frac{n-1}{n}.
\end{equation}
The minimum distance of the SBCs conceived is $d = 2$, hence this guarantees the convergence of iterative decoding to a vanishingly low BER, when they constitute the outer code~\cite{benedetto1998serial, land2005reliability}.

By exploiting the classical-to-quantum isomorphism~\cite{chandra2017quantum1}, this specific type of classical SBCs can be readily transformed into their quantum counterparts. To elaborate a little further, given a pair of classical codes $\mathcal{C}_1(n, k_1, d_1)$ and $\mathcal{C}_2(n, k_2, d_2)$ having PCMs $\mathbf{H}_1$ and $\mathbf{H}_2$, respectively, a QSC $\mathcal{C}[\tilde{n}, \tilde{k}, \tilde{d}]$ having a binary PCM $\mathbf{H}$ can be constructed from $\mathbf{H}_x = \mathbf{H}_1$ and $\mathbf{H}_z = \mathbf{H}_2$ so that $\mathbf{H}_x$ will be used for mitigating the bit-flip errors and $\mathbf{H}_z$ will be used for mitigating the phase-flip errors, where we have $\tilde{n} = n$, $\tilde{k} = k_1 + k_2 - n$, and $\tilde{d} = \min(d_1,d_2)$. In general, there are two ways of constructing the binary PCM $\mathbf{H}$ of a QSC $\mathcal{C}$ given a pair of PCMs $\mathbf{H}_x$ and $\mathbf{H}_z$. Firstly, we may construct a Calderbank-Steane-Shor (CSS) type quantum code, whose binary PCM $\mathbf{H}$ is as follows~\cite{calderbank1996good}:
\begin{equation}
\mathbf{H} = \left[ \begin{array}{c|c} \mathbf{H}_z & \mathbf{0}  \\ \mathbf{0} & \mathbf{H}_x  \end{array} \right].
\end{equation}
Secondly, we may also construct a non-CSS type quantum code, whose binary PCM $\mathbf{H}$ is given by~\cite{calderbank1997quantum}
\begin{equation}
\mathbf{H} = \left[ \begin{array}{c|c} \mathbf{H}_z & \mathbf{H}_x \end{array} \right].
\end{equation}
In order to conceive a valid PCM $\mathbf{H}$ for a QSC $\mathcal{C}$, a pair of PCMs $\mathbf{H}_x$ and $\mathbf{H}_z$ has to satisfy the symplectic criterion, which is defined as~\cite{calderbank1997quantum}
\begin{equation}
\mathbf{H}_z.\mathbf{H}_x^T + \mathbf{H}_x^T.\mathbf{H}_z = 0.
\end{equation}
Therefore, the symplectic criterion formulated for a CSS type quantum code can be reduced to
\begin{equation}
\mathbf{H}_z.\mathbf{H}_x^T = 0.
\label{eq:symplectic}
\end{equation}
A specific case of CSS type quantum codes, where we have $\mathbf{H}_x = \mathbf{H}_z$, is referred to as a dual-containing CSS quantum code. Therefore, the dual-containing CSS quantum codes can be instantly derived from the classical SBCs having the following PCMs:
\begin{equation}
\mathbf{H}_z^T = \mathbf{H}_x^T = \mathbf{J}_{1,n},
\end{equation}
where $n$ is an even number. This automatically satisfies the symplectic criterion of Eq.~\eqref{eq:symplectic}. Ultimately, the resultant PCM of QSBCs is given by:
\begin{equation}
\mathbf{H} = \left[ \begin{array}{c|c} \mathbf{J}_{1,n} & \mathbf{0}  \\ \mathbf{0} & \mathbf{J}_{1,n}  \end{array} \right].
\end{equation}
Hence, for dual-containing CSS quantum codes, the relationship between the classical coding rate $r_C$ and the quantum coding rate $r_Q$ can be described as follows~\cite{chandra2018quantum2, babar2015road, babar2018duality}
\begin{equation}
r_Q = 2r_C -1. 
\end{equation} 
Since the quantum coding rate $r_Q$ has to be positive $(r_Q > 0)$, the original classical code must exhibit a classical coding rate of $r_C > 1/2$.

\begin{figure}[ht!]
\center
\includegraphics[width=\linewidth]{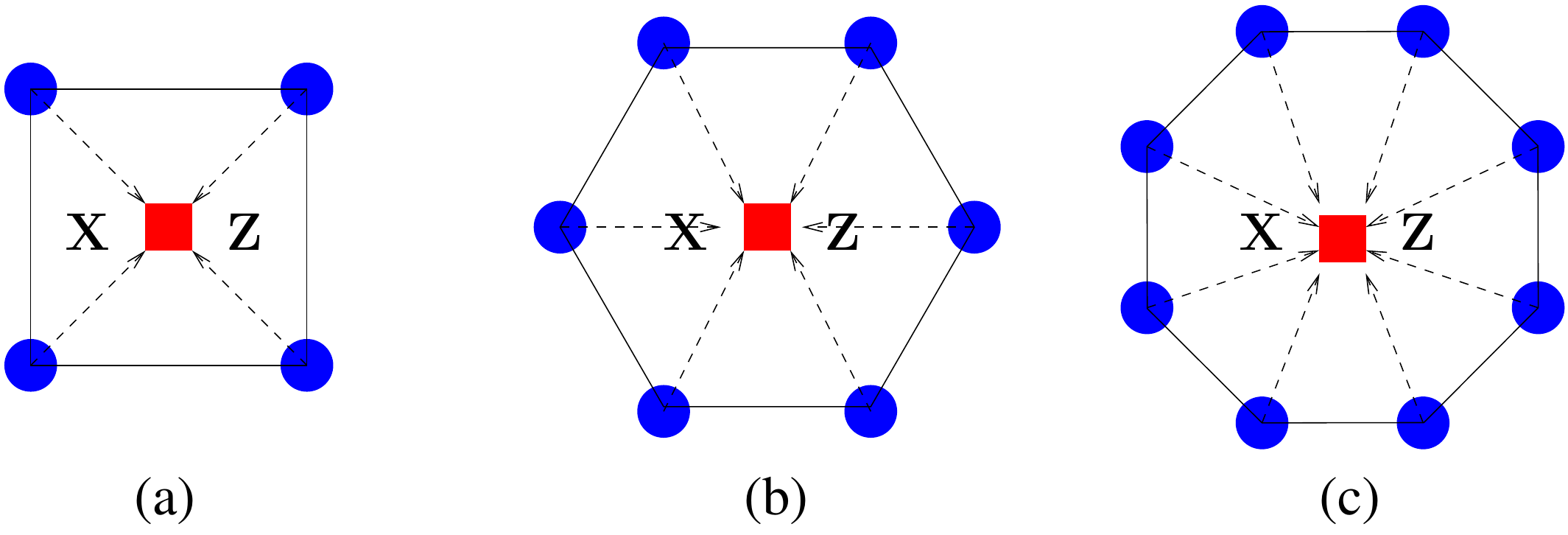}
\caption{QSBCs can be arranged to match the qubit layout on a lattice structure. The blue circles represent the physical qubits, while the red circles denote the stabilizer operators. Each red square is utilized for measuring both $\mathbf{X}$ and $\mathbf{Z}$ operator. Hence, the resultant quantum codes belong to the family of dual-containing CSS codes. The quantum coding rate $r_Q$ for construction (a) is $1/2$, for (b) is $2/3$, and for (c) is $3/4$. All of the QSBCs exhibit a minimum distance of $d = 2$.}
\label{fig:topology}
\end{figure}

For QSCs, the PCM $\mathbf{H}$ is associated with stabilizer operators $S_i \in \mathcal{S}$. For example, let us consider a classical SBC $\mathcal{C}(4,3,2)$ having a PCM of 
\begin{equation}
\mathbf{H}_x = \mathbf{H}_z = [1 \ 1 \ 1 \ 1],
\end{equation}
exhibiting a classical coding rate of $r_C = 3/4$, which is associated with a QSBC of $\mathcal{C}[n,k,d] = \mathcal{C}[4,2,2]$ having a quantum coding rate of $r_Q = 1/2$. The stabilizer operators of $\mathcal{C}[4,2,2]$ are given by\footnote{The shortened representation of stabilizer operators is used for simplifying the original representation of stabilizer operators. For example, the sorthened version of $S_1 = \mathbf{Z}_1\mathbf{Z}_2\mathbf{Z}_3\mathbf{Z}_4$ is used for simplifying $S_1 = \mathbf{Z}_1 \otimes \mathbf{Z}_2 \otimes \mathbf{Z}_3 \otimes \mathbf{Z}_4$, where the notation $\otimes$ represents a tensor product. For the rest of the paper, we always use the shortened representation of stabilizer operators.}
\begin{align}
S_1 & = \mathbf{Z}_1\mathbf{Z}_2\mathbf{Z}_3\mathbf{Z}_4, \nonumber \\
S_2 & = \mathbf{X}_1\mathbf{X}_2\mathbf{X}_3\mathbf{X}_4,
\end{align}
where $\mathbf{X}$ and $\mathbf{Z}$ are the Pauli matrices. By exploiting the classical-to-quantum isomorphism, we arrive at the PCM $\mathbf{H}$ of the $\mathcal{C}[4,2,2]$ dual-containing CSS quantum code formulated as
\begin{equation}
\mathbf{H}_{[4,2,2]} = \left[ \begin{array}{cccc|cccc} 1 & 1 & 1 & 1 & 0 & 0 & 0 & 0 \\ 0 & 0 & 0 & 0 & 1 & 1 & 1 & 1 \end{array} \right].
\end{equation}
Similarly, we can readily extend the construction to a higher quantum coding rate, as exemplified by $\mathcal{C}[n,k,d] = \mathcal{C}[6,4,2]$, which is derived from a classical SBC of $\mathcal{C}(6,5,2)$. Hence, the stabilizer operators for $\mathcal{C}[6,4,2]$ are defined by
\begin{align}
S_1 & = \mathbf{Z}_1\mathbf{Z}_2\mathbf{Z}_3\mathbf{Z}_4\mathbf{Z}_5\mathbf{Z}_6, \nonumber \\
S_2 & = \mathbf{X}_1\mathbf{X}_2\mathbf{X}_3\mathbf{X}_4\mathbf{X}_5\mathbf{X}_6.
\end{align}
For this construction, the resultant quantum coding rate is $r_Q = 2/3$. The same analogy can be used for constructing $\mathcal{C}[n,k,d] = \mathcal{C}[8,6,2]$ derived from a classical SBC $\mathcal{C}(8,7,2)$. Therefore, the stabiizer operators are as follows:
\begin{align}
S_1 & = \mathbf{Z}_1\mathbf{Z}_2\mathbf{Z}_3\mathbf{Z}_4\mathbf{Z}_5\mathbf{Z}_6\mathbf{Z}_7\mathbf{Z}_8, \nonumber \\
S_2 & = \mathbf{X}_1\mathbf{X}_2\mathbf{X}_3\mathbf{X}_4\mathbf{X}_5\mathbf{X}_6\mathbf{X}_7\mathbf{X}_8.
\end{align}
The resultant QSBC exhibits a quantum coding rate of $r_Q = 3/4$. The discussion on how the stabilizer operators $S_i$ can be invoked for detecting quantum errors will be discussed in subsection~\ref{Stabilizer Measurement for QSBCs}.

Furthermore, the QSBCs can also be classified as a family of QTECCs, as shown in Fig.~\ref{fig:topology}. Assuming that we can arrange the physical qubits on the vertices of a lattice structure, it inherently provides a localized stabilizer measurement property. Since the resultant QSC constructions are only capable of error detection, which is a consequnce of having a minimum distance of $d = 2$, the stabilizer measurements can only indicate the presence or the absence of quantum errors, but not the specific numbers or the position of the errors. However, we will show later that this error detection capability can be transformed into error correction capability by concatenating QSBCs with a carefully selected inner code.

\subsection{Quantum Encoder}
\label{Quantum Encoder}

In quantum domain, a $k$-logical qubit information word in the state of $|\psi\rangle$ can be transformed into a $n$-physical qubit codeword in the state of $|\overline{\psi}\rangle$, where $n > k$, with the aid of $(n-k)$ auxiliary qubits initialized in the state of $|0\rangle$. This specific transformation, which is carried out by a unitary transformation $\mathcal{V}$ referred to as a quantum encoder, can be formally described as follows\footnote{The superscript of $n$ in notation $|\psi\rangle^n$ denotes the number of physical qubits, which is $n$, given a quantum state $|\psi\rangle$. This notation will be used throughout this treatise.}:
\begin{equation}
\mathcal{V} \left( |\psi\rangle^k \otimes |0\rangle^{\otimes (n-k)} \right) = |\overline{\psi}\rangle^n.
\end{equation}
This process is reminiscent of the encoding process of classical error correction codes. To elaborate, in the classical domain, we can transform a $k$-bit information word into an $n$-bit codeword with the aid of the generator matrix $\mathbf{G}$, where the additional of $(n-k)$ bits are referred to as the redundant bits.  

Based on the description of QSBCs in Subsection~\ref{Codes Construction}, they can be classified as a member of the dual-containing CSS code family. For this specific class of quantum codes, the design of the quantum encoder $\mathcal{V}$ can be readily derived from its classical PCMs~\cite{botsinis2016quantum, mackay2004sparse}. More specifically, the PCM of the classical code $\mathcal{C}(n,k,d)$ may be utilized to obtain the quantum encoder $\mathcal{V}$ of a QSC $\mathcal{C}[\tilde{n},\tilde{k},\tilde{d}]$, where $\tilde{n} = n$, $\tilde{k} = 2k-n$, and $\tilde{d} = d$. 

Let us now embark on creating the quantum encoder $\mathcal{V}$ of dual-containing CSS codes derived from the classical codes $\mathcal{C}(n,k,d)$. The PCM of the classical code $\mathcal{C}(n,k,d)$ can be represented by a full-rank matrix $\mathbf{H}$ having $n \times (n-k)$ elements. Naturally, every PCM $\mathbf{H}$ of linear block codes can be transformed into the corresponding systematic form of
\begin{equation}
\widetilde{\mathbf{H}} = \left[ \mathbf{I}_{n-k} | \mathbf{A}_{n-k,k} \right]
\end{equation}
by using row operations and column permutations, where $\mathbf{I}_{n-k}$ is a $(n-k)$-dimension identity matrix, and the matrix $\mathbf{A}$ has $(n-k) \times k$ elements. For the next step, we may further reduce the matrix $\mathbf{A}$ into another systematic form of
\begin{equation}
\widetilde{\mathbf{A}} = \left[ \mathbf{I}_{n-k} | \mathbf{B}_{n-k,2k-n} \right],
\end{equation}
where $\mathbf{I}_{n-k}$ is another $(n-k)$-dimension identity matrix and the matrix $\mathbf{B}$ has $(n-k) \times (2k-n)$ elements.

Ultimately, the quantum encoder $\mathcal{V}$ of a dual-containing CSS code can be described as a two-stage encoder. The first stage of the quantum encoder is used for initializing a set of codewords $\mathcal{C}$, which must not have a difference exactly corresponding to a specific legitimate codewords of $\mathcal{C}^{\perp}$, where $\mathcal{C}^{\perp}$ is the dual code of $\mathcal{C}$. Therefore, it can be utilized for generating the unique cosets of $\mathcal{C}$ relative to $\mathcal{C}^{\perp}$. Next, the second stage of quantum encoder $\mathcal{V}$ is invoked for generating the code space of $\mathcal{C}^{\perp}$ according to the PCM $\widetilde{\mathbf{H}}$. Hence, the resultant states after the first and the second stage are constituted by the superposition of all the codewords of $\mathcal{C}^{\perp}$ generated by the second stage added to the codewords of $\mathcal{C}$ generated in the first stage. The more general method of constructing the quantum encoder $\mathcal{V}$ for all types of QSCs, including the non-dual-containing CSS codes and non-CSS codes, can be found in~\cite{cleve1997efficient, djordjevic2012quantum}.

\begin{figure}[ht!]
\center
\includegraphics[width=\linewidth]{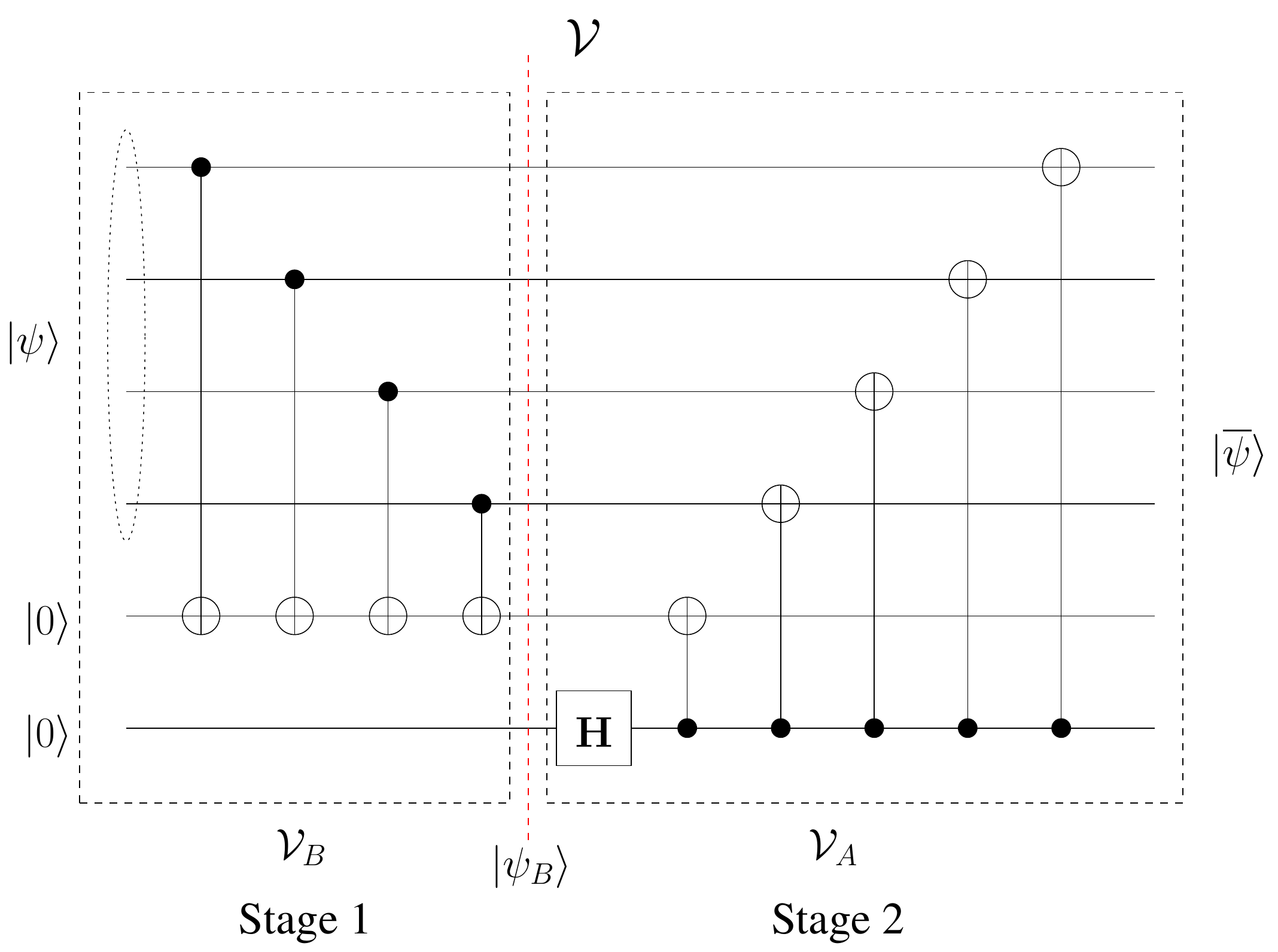}
\caption{The quantum encoder $\mathcal{V}$ can be described as a two-stage encoder. For the QSBCs, the CNOT connections of the first stage $\mathcal{V}_B$ are defined by matrix $\mathbf{B}$ of Eq.~\eqref{eq:matrix_B}, while those of the second stage $\mathcal{V}_A$ are defined by matrix $\mathbf{A}$ of Eq.~\eqref{eq:matrix_A}.}
\label{fig:stage}
\end{figure}

For gleaning a clearer idea about the two-stage quantum encoder $\mathcal{V}$, let us consider the QSBC $\mathcal{C}\left[6,4,2\right]$ and construct its quantum encoder $\mathcal{V}$ based on the classical PCM $\mathbf{H}$ derived from a classical code $\mathcal{C}(6,5,2)$ as follows: 
\begin{equation}
\mathbf{H} = [1 \ 1 \ 1 \ 1 \ 1 \ 1].
\label{eq:matrix_H}
\end{equation}
Fortunately, the PCM $\mathbf{H}$ has already a systematic structure, hence we have $\mathbf{H} = \widetilde{\mathbf{H}}$. From the PCM $\mathbf{H}$ in Eq.~\eqref{eq:matrix_H}, matrix $\mathbf{A}$ is readily given by
\begin{equation}
\mathbf{A} = [1 \ 1 \ 1 \ 1 \ 1].
\label{eq:matrix_A}
\end{equation} 
Consequently, given that $\mathbf{A} = \widetilde{\mathbf{A}}$, we obtain the matrix $\mathbf{B}$ as follows:
\begin{equation}
\mathbf{B} = [1 \ 1 \ 1 \ 1].
\label{eq:matrix_B}
\end{equation}

\begin{figure*}[ht!]
\centering
\begin{subfigure}{0.40\linewidth}
\includegraphics[width=\linewidth]{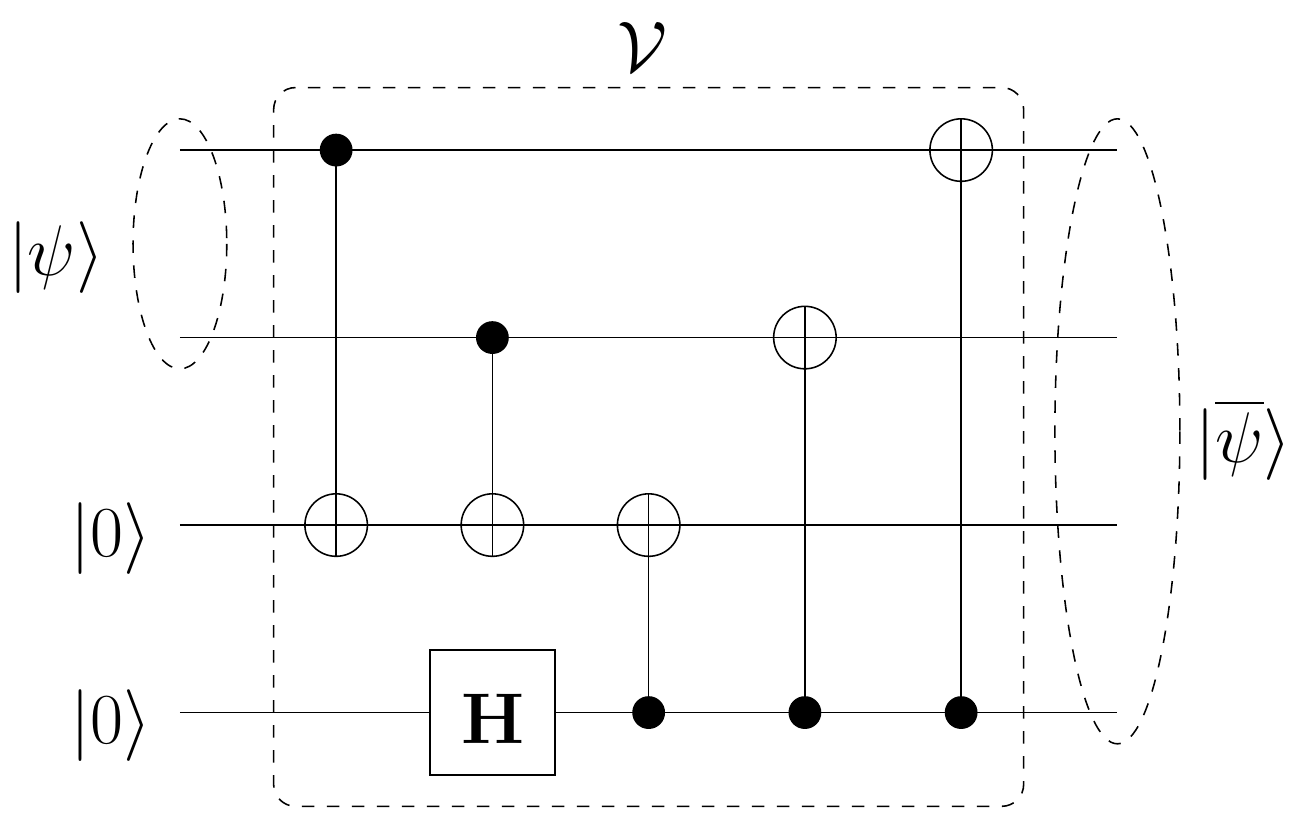}
\caption{The quantum encoder $\mathcal{V}$ for $\mathcal{C}[4,2,2]$.}
\label{fig:circuit_layout_1}
\vspace{5mm}
\end{subfigure}
\begin{subfigure}{0.50\linewidth}
\includegraphics[width=\linewidth]{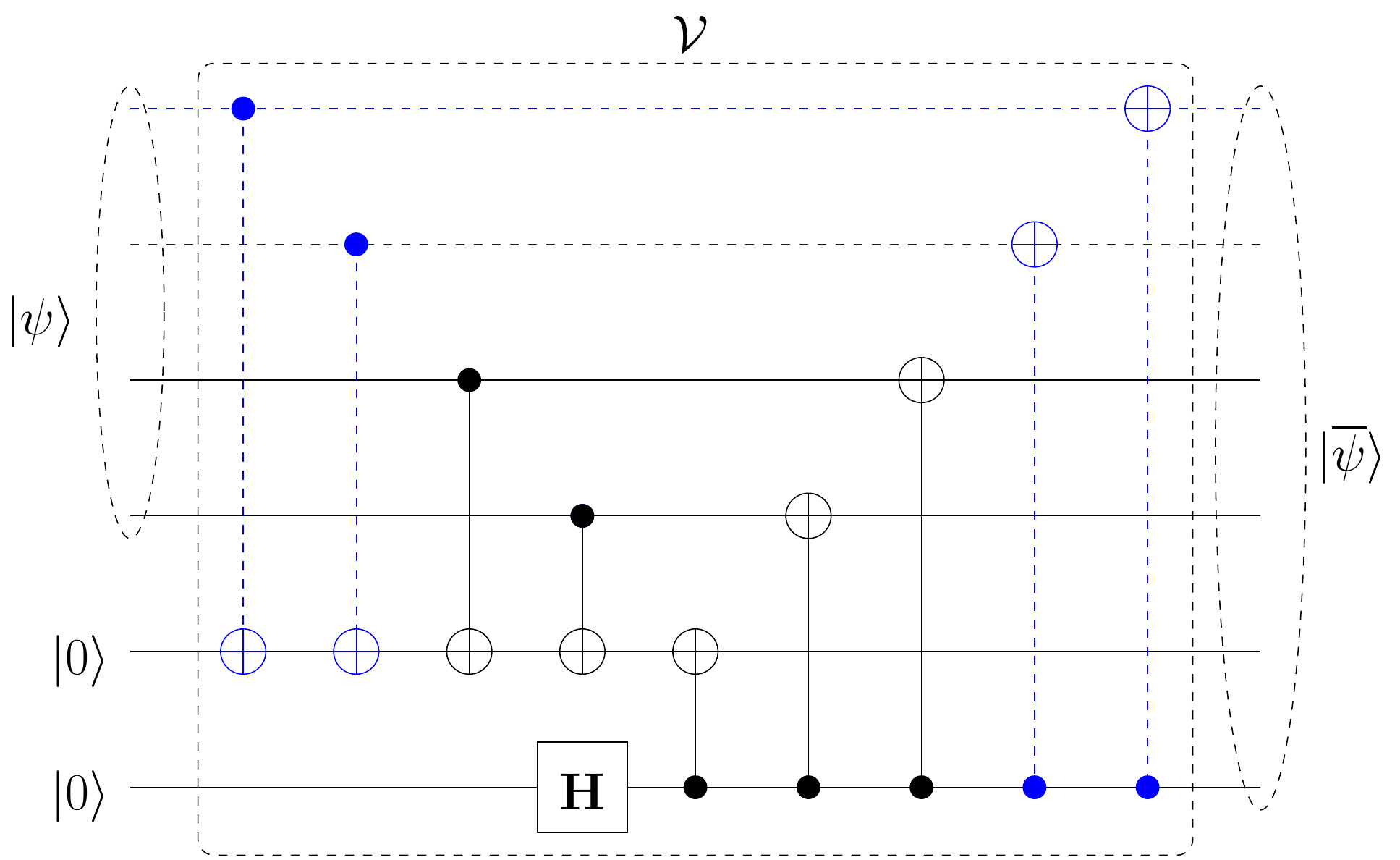}
\caption{The quantum encoder $\mathcal{V}$ for $\mathcal{C}[6,4,2]$.}
\label{fig:circuit_layout_2}
\vspace{5mm}
\end{subfigure}
\begin{subfigure}{0.60\linewidth}
\includegraphics[width=\linewidth]{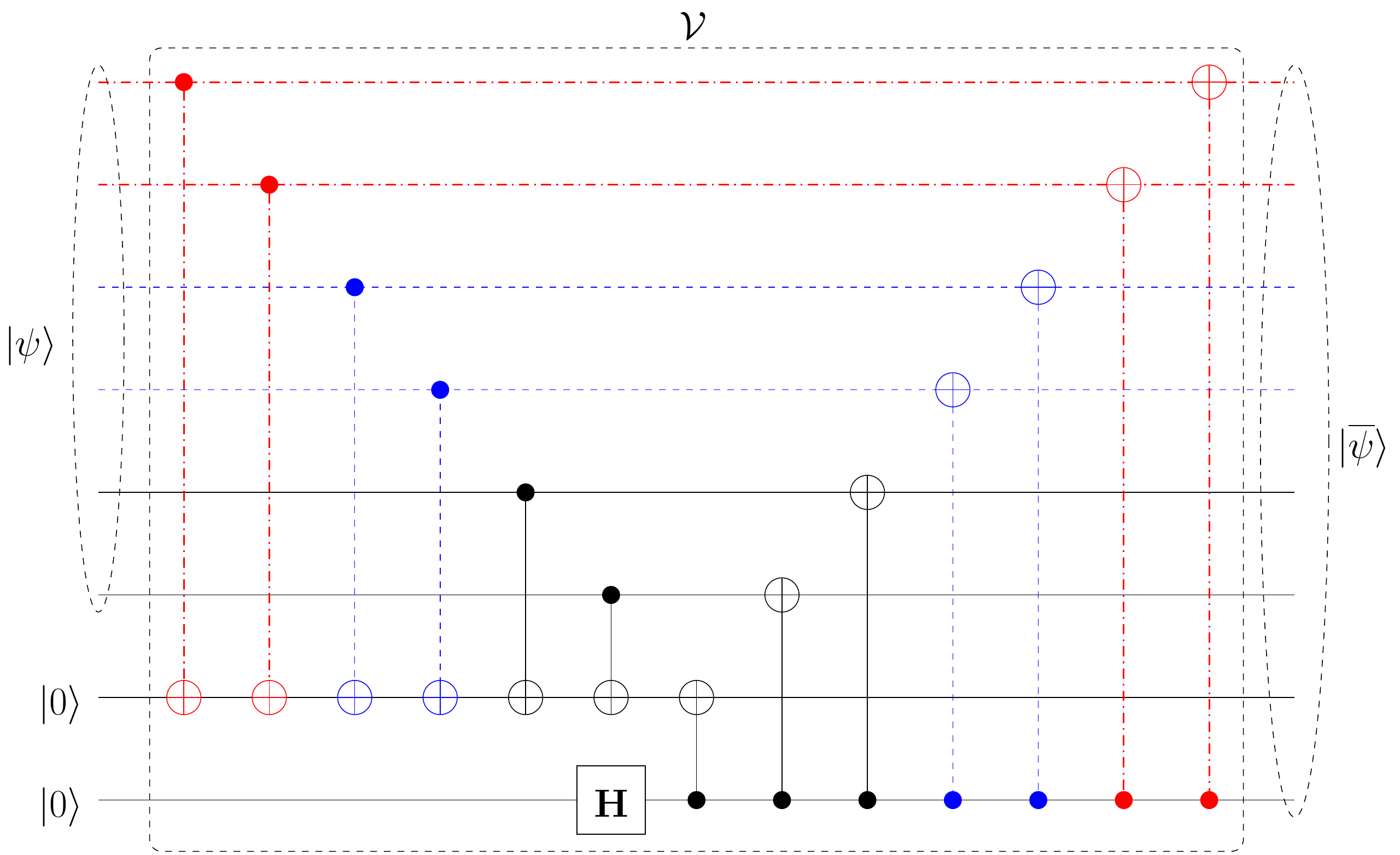}
\caption{The quantum encoder $\mathcal{V}$ for $\mathcal{C}[8,6,2]$.}
\label{fig:circuit_layout_3}
\vspace{5mm}
\end{subfigure}
\caption{The QSBC encoders $\mathcal{V}$ for various quantum coding rates illustrating the flexibility and scalability of the code constructions. Figure (a) depicts the quantum encoder $\mathcal{V}$ of the QSBC $\mathcal{C}[4,2,2]$. By incorporating additional gates and connections denoted by the dashed blue lines in Fig. (b) into the quantum encoder (a), we obtain the quantum encoder $\mathcal{V}$ of the QSBC $\mathcal{C}[6,4,2]$. Similarly, by adding more gates and connections denoted by a dashed-dotted red lines to the quantum encoder (b), we can readily create the quantum encoder $\mathcal{V}$ of the QSBC $\mathcal{C}[8,6,2]$. In other words, we can construct the quantum encoder $\mathcal{V}$ of QSBCs for a high quantum coding rate, which will simultaneously contain the quantum encoder $\mathcal{V}$ of the lower quantum coding rate. Therefore, it is no longer necessary to build more than one quantum encoder $\mathcal{V}$ for various quantum coding rate QSBCs because the quantum encoder $\mathcal{V}$ of QSBCs exhibit a self-contained structure. This is reminiscent of the classical rate-compatible punctured codes proposed in~\cite{hagenauer1988rate}.}
\label{fig:circuit_layout}
\end{figure*}

Based on Eq.~\eqref{eq:matrix_A} and~\eqref{eq:matrix_B}, the first stage and the second stage of the quantum encoder $\mathcal{V}$ of the QSBC $\mathcal{C}[6,4,2]$ is denoted by $\mathcal{V}_B$ and $\mathcal{V}_A$, respectively, in Fig.~\ref{fig:stage}. To elaboreate a little further, the CNOT connections between the logical qubits in the state of $|\psi\rangle$ and the first auxiliary qubits are defined by the matrix $\mathbf{B}$. More specifically, given the element $b_{i,j}$ of the matrix $\mathbf{B}$, if the value of $b_{i,j} = 1$, it means that the $j$-th logical qubit controls the CNOT connection of the $i$-th auxiliary target qubit. As an example, based on the matrix $\mathbf{B}$ in Eq.~\eqref{eq:matrix_B}, we can see that at the first stage $\mathcal{V}_B$ of the quantum encoder $\mathcal{V}$ the first auxiliary qubit is controlled by all four logical qubits in the state of $|\psi\rangle$, since the matrix $\mathbf{B}$ contains all 1 elements.

Consequently, the first stage $\mathcal{V}_B$ of the quantum encoder $\mathcal{V}$ transforms the first $m$ auxiliary qubits, which are initialized to the state of $|0\rangle$, according to the logical qubits $|\psi\rangle$. Explicitly, given that $|\psi\rangle = |\mathbf{c}\rangle$, where $c$ is a $k$-bit binary string, the first stage $\mathcal{V}_B$ of the quantum encoder $\mathcal{V}$ transforms the auxiliary qubits into the state of $|\mathbf{Bc}\rangle$. Therefore, the state of physical qubits at the output of the first stage $\mathcal{V}_B$, namely $|\psi_B\rangle$, created by the action of the first stage $\mathcal{V}_B$ of quantum encoder $\mathcal{V}$ of Fig.~\ref{fig:stage} can be expressed as follows:
\begin{equation}
|\psi_B\rangle = \mathcal{V}_B\left( |\mathbf{c}\rangle^{k} |\mathbf{0}\rangle^{\otimes m} |\mathbf{0}\rangle^{\otimes m} \right) = |\mathbf{c}\rangle^{k} |\mathbf{B}\mathbf{c}\rangle^{m} |\mathbf{0}\rangle^{\otimes m},
\label{eq:first_stage}
\end{equation}
where $m = n-k$. Hence, as we have mentioned earlier that the first stage $\mathcal{V}_B$ of quantum encoder $\mathcal{V}$ creates a set of codewords $\mathcal{C}$, which must not have a difference exactly corresponding to a specific legitimate codeword in $\mathcal{C}^{\perp}$. For example, based on the PCM $\mathbf{H}$ of a classical code $\mathcal{C}(6,5,2)$ given in Eq.~\eqref{eq:matrix_H}, we can construct the code space of the dual code $\mathcal{C}^{\perp}$ as follows:
\begin{equation}
\mathcal{C}^{\perp} = \lbrace 000000, 111111 \rbrace.
\end{equation}
For a given $k$-bit binary string of $\mathbf{c}$, the binary string of $\left[ \mathbf{c}, \mathbf{Bc}, \mathbf{0} \right]$ is indeed a codeword in $\mathcal{C}$. Let us denote the set of all the $n$-bit binary string of $\left[ \mathbf{c}, \mathbf{Bc}, \mathbf{0} \right]$ as $\mathcal{B}$, which basically creates a subspace of $\mathcal{C}$, i.e. $\mathcal{B} \subseteq \mathcal{C}$. More explicitly, the code space of $\mathcal{B}$ based on matrix $\mathbf{B}$ of Eq.~\eqref{eq:matrix_B} is given as follows:
\begin{align}
\mathcal{B} = \lbrace & 000000, 000110, 001010, 001100, \nonumber \\
& 010010, 010100, 011000, 011110, \nonumber \\
& 100010, 100100, 101000, 101110, \nonumber \\
& 110000, 110110, 111010, 111100 \rbrace.
\end{align}
It is clear that none of the codeword in $\mathcal{B}$ differs by one element in $\mathcal{C}^{\perp}$, i.e. adding any codeword from the non-zero codeword in $\mathcal{C}^{\perp}$ to the one of the codeword in $\mathcal{B}$ does not yield another codeword in $\mathcal{B}$, since the last $m$ bits of $\mathcal{B}$ contains all 0 element, while the non-zero codeword of $\mathcal{C}^{\perp}$ has 1s in the last $m$ bits. Hence, each $k$-bit binary string of $\mathbf{c}$ creates a unique coset in $\mathcal{C}$ relative to the $\mathcal{C}^{\perp}$.

The second stage $\mathcal{V}_A$ of the quantum encoder $\mathcal{V}$ is started by initializing the remaining $m = n - k$ auxiliary qubits in the state of $|+\rangle$ states by using $m$ Hadamard gates, which can be expressed mathematically as follows:
\begin{align}
\mathbf{H}^{\otimes m}|\mathbf{0}\rangle^{\otimes m} = |+\rangle^{\otimes m} & = \left( \frac{|0\rangle + |1\rangle}{\sqrt{2}}\right)^{\otimes m} \nonumber \\
 & = \frac{1}{2^{m/2}}\sum^{2^m -1}_{i}|\mathbf{i}\rangle.
 \label{eq:equalsuperposition}
\end{align}
Explicitly, Eq.~\eqref{eq:equalsuperposition} basically represents a sum of an equal superposition of all possible states over all $2^m$ binary strings of length $m$. Next, we combine the initialized equal superposition of Eq.~\eqref{eq:equalsuperposition} with the output of the first stage $|\psi_B\rangle$. Let us assume that string $\mathbf{t}$ denotes the string of $[\mathbf{c},\mathbf{Bc}]$. The effect of this operation is to add rows of PCM $\widetilde{\mathbf{H}} = [\mathbf{I}_{n-k}|\mathbf{A}_{n-k,k}]$ to the binary string $\mathbf{t}$. Hence, for a given $k$-bit binary input string of $\mathbf{c}$, the final state of the physical qubits after the first and the second stage of the quantum encoder $\mathcal{V}$ can be expressed as
\begin{equation}
|\overline{\psi}\rangle = \frac{1}{2^{m/2}} \sum_{\textbf{r} \in \mathcal{C}^{\perp}(\mathbf{H})} |\mathbf{r + t}\rangle^{n}.
\label{eq:second_stage}
\end{equation} 
Finally, if the state of the $k$ logical qubits is expressed in the form of the superposition the binary strings $\mathbf{c}_i$ as follows:
\begin{equation}
|\psi\rangle^k = \sum_{i}^{2^k - 1} p_i |\mathbf{c}_i\rangle^k,
\end{equation}
then the output state of the physical qubits $|\overline{\psi}\rangle$ can be formulated as
\begin{equation}
|\overline{\psi}\rangle^n = \sum_{\mathbf{y} \in \mathcal{C}(\mathbf{H})} p_{\mathbf{y}} \sum_{\mathbf{r} \in \mathcal{C}^{\perp}(\mathbf{H})}|\mathbf{r} + \mathbf{y}\rangle^n.
\end{equation}
Readers who might be interested in different examples of creating encoders for various dual-containing CSS codes exemplified by Steane's code $\mathcal{C}[7,1,3]$ of~\cite{steane1996multiple} and by the quantum Bose-Chaudhuri-Hocquenghem (QBCH) code $\mathcal{C}[15,7,3]$ of~\cite{grassl1999quantumbch}, please refer to~\cite{botsinis2016quantum}.

The resultant quantum encoders $\mathcal{V}$ conceived for the QSBCs $\mathcal{C}[4,2,2]$, $\mathcal{C}[6,4,2]$, and $\mathcal{C}[8,6,2]$ can be seen in Fig.~\ref{fig:circuit_layout}, where it shows that the QSBC encoders $\mathcal{V}$ exhibit the natural design for flexibility. The inverse encoder $\mathcal{V}^{\dagger}$ is implemented with the aid of the same exact quantum circuit, where the input and output qubits are reversed.

\subsection{Stabilizer Measurement for QSBCs}
\label{Stabilizer Measurement for QSBCs}

The QSCs are capable of predicting both the number and the position of errors without actually observing the states of the physical qubits. In order to achieve this, a syndrome-decoding-like method was introduced~\cite{gottesman1997stabilizer}. Instead of observing the information within the physical qubits, which would collapse the superposition state to a classical state, a set of auxiliary qubits are prepared for observing the syndrome of the physical qubits using the so-called stabilizer operators. A stabilizer operator $S_i$ belonging to the stabilizer group $\mathcal{S} \in \mathcal{P}_n$ is an $n$-tuple Pauli operator, which stabilizes the state of the encoded physical qubits $|\overline{\psi}\rangle$ may be formulated as follows:
\begin{equation}
S_i|\overline{\psi}\rangle = |\overline{\psi}\rangle.
\end{equation}

\begin{figure*}[ht!]
\centering
\begin{subfigure}{0.75\linewidth}
\includegraphics[width=\linewidth]{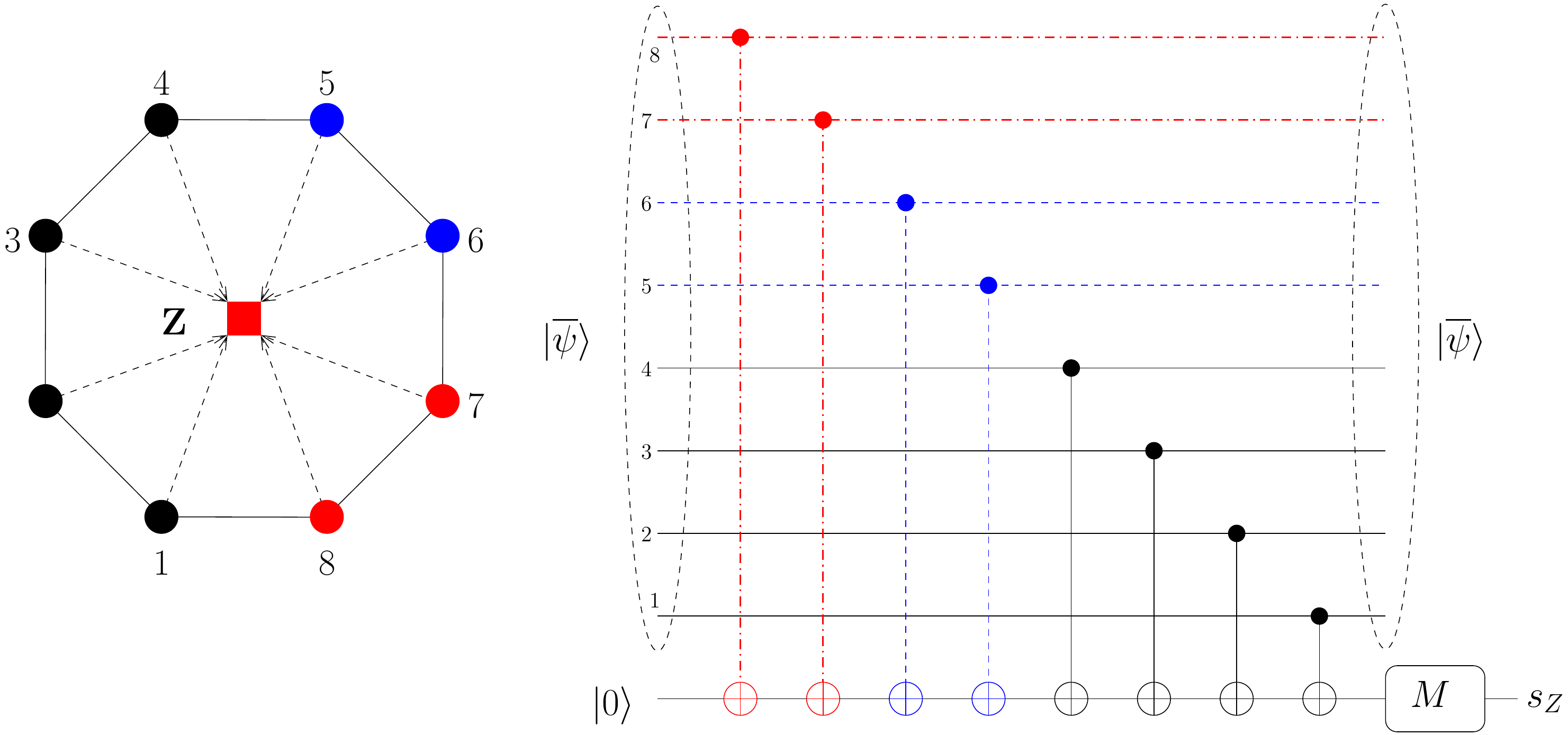}
\caption{The $\mathbf{Z}$-stabilizer measurements for QSBCs.}
\label{fig:syndromez}
\vspace{5mm}
\end{subfigure}
\begin{subfigure}{0.75\linewidth}
\includegraphics[width=\linewidth]{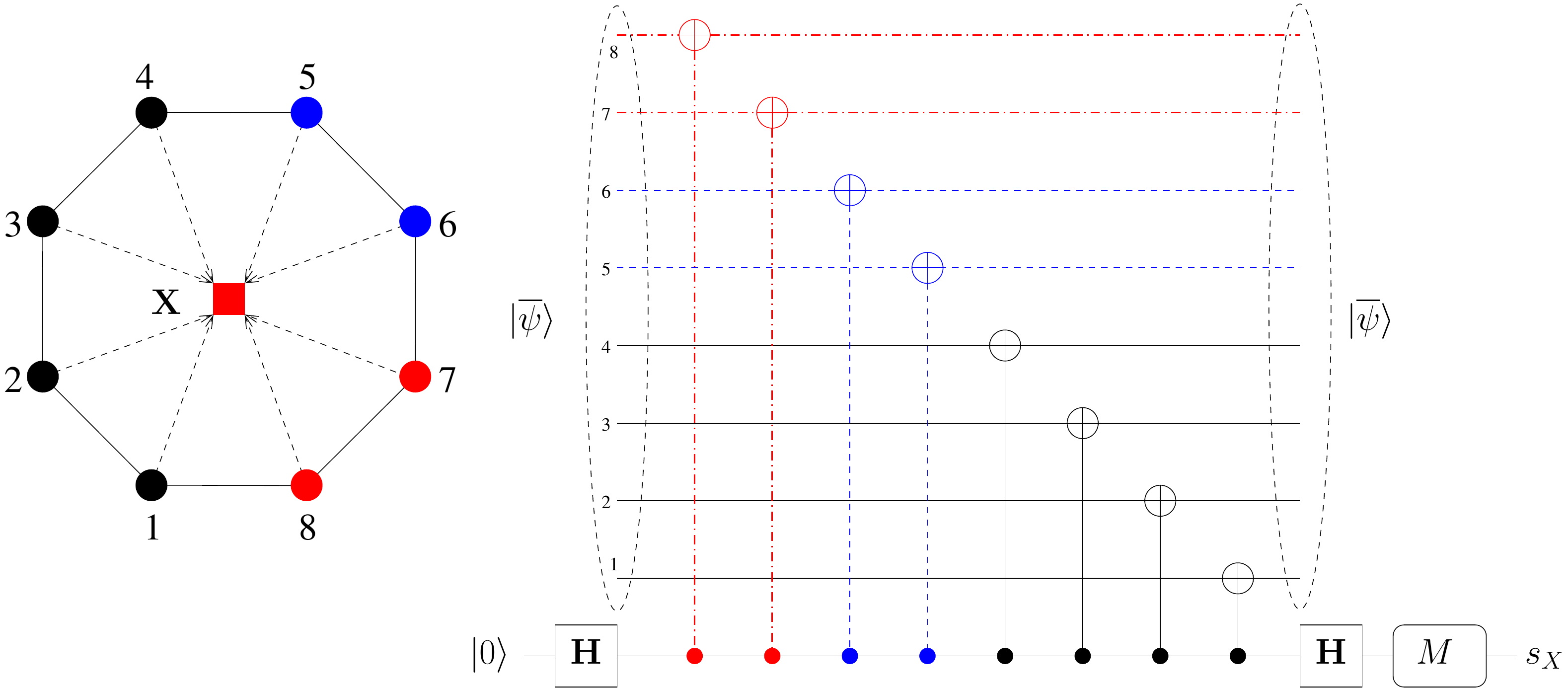}
\caption{The $\mathbf{X}$-stabilizer measurement for QSBCs.}
\label{fig:syndromex}
\vspace{5mm}
\end{subfigure}
\caption{The stabilizer measurements of QSBCs for various quantum coding rates of QSBCs illustrating the flexibility and scalability of the encoder constructions. Even though the QSBCs can be defined as illustrated in Fig.~\ref{fig:topology}, from implementation perspective the QSBCs can be arranged to be more scalable. For example, all the physical qubits for QSBCs having quantum coding rate of $r_Q = \lbrace 1/2, 2/3, 3/4\rbrace$ can be arranged on an octagon.}
\label{fig:stabilizer}
\end{figure*}

An error operator $P \in \mathcal{P}_n$ inflicted by the quantum channel on the encoded state of physical qubits transforms the legitimate state $|\overline{\psi}\rangle$ into the contaminated received state $|\widetilde{\psi}\rangle$, which can be expressed as follows: 
\begin{equation}
|\widetilde{\psi}\rangle = {P}|\overline{\psi}\rangle.
\end{equation}
The syndrome values can be obtained by performing eigenvalue-based measurement of the received state $|\widehat{\psi}\rangle$ assisted by the auxiliary qubits, which can be defined as follows:
\begin{equation}
S_i|\widetilde{\psi}\rangle = \left\{ \begin{array}{lll} |\widetilde{\psi}\rangle & , & S_i{P} = {P}S_i  \\ -|\widetilde{\psi}\rangle & , & S_i{P} = -{P}S_i. \end{array} \right.
\end{equation}
The $\pm 1$ eigenvalues attained from the stabilizer measurement act similarly to the $\lbrace 0, 1 \rbrace$ values of the classical syndrome measurements. Hence, they also can be used for inferring both the number and the position of errors without actually observing the state of the physical qubits. 

As we briefly discussed, the eigenvalue measurements require an extra auxiliary qubit for each stabilizer measurement. As for QSBCs, the stabilizer measurement can be implemented using the circuits seen in Fig.~\ref{fig:stabilizer}. More specifically, Fig.~\ref{fig:stabilizer}(a) depicts the $\mathbf{Z}$-stabilizer measurement, while Fig~\ref{fig:stabilizer}(b) portrays the $\mathbf{X}$-stabilizer measurement. It can be clearly observed from Fig.~\ref{fig:stabilizer} that for both $\mathbf{X}$ and $\mathbf{Z}$ stabilizer measurements, the circuit constructed for realizing the stabilizer measurements of a QSBC having a higher quantum coding rate inherently contains the stabilizer measurements required for a QSBC having a lower quantum coding rate. To elaborate a little further, in Fig.~\ref{fig:stabilizer}, the circuit implementation of the stabilizer measurements of a $1/2$-rate QSBC is highlighted using black solid lines. In case we want to employ another QSBC having a higher quantum coding rate, for example a $2/3$-rate QSBC, we can simply incorporate the stabilizer measurement from the $1/2$-rate scheme and add further gates, which are highlighted using blue dashed lines in Fig.~\ref{fig:stabilizer}, without changing the stabilizer measurement circuit. A similar approach is applicable when we want to employ a $3/4$-rate QSBC. We incorporate the stabilizer measurements of the $2/3$-rate QSBC and then add more gates, which are highlighted using red dashed lines. Ultimately, we have shown that the nature of the PCMs from the QSBCs leads to a very convenient design for their quantum encoders and for their stabilizer measurements, which are capable of supporting multiple quantum coding rates of $r_Q = \frac{k}{k+1}$ using a single quantum circuit implementation. 

\subsection{Classical Simulation for QSBCs}
\label{Classical Simulation for QSBCs}

The quantum encoder $\mathcal{V}$ of a QSBC and its inverse encoder $\mathcal{V}^{\dagger}$ are composed of quantum Clifford gates. This implies that they can be conveniently simulated using classical computers~\cite{gottesman1998heisenberg}. An $n$-tuple Pauli operator $P \in \mathcal{P}_n$ can be represented by a $2n$-element binary vector, where each of the $n$-element binary vectors constitutes a Pauli $\mathbf{Z}$ and a Pauli $\mathbf{X}$ component. The mapping of the Pauli matrix to the associated binary vector can be formulated as follows:
\begin{align}
\mathbf{I} & \rightarrow \left[ \begin{array}{c|c} 0 & 0\end{array}\right], \nonumber \\
\mathbf{X} & \rightarrow \left[ \begin{array}{c|c} 0 & 1\end{array}\right], \nonumber \\
\mathbf{Y} & \rightarrow \left[ \begin{array}{c|c} 1 & 1\end{array}\right], \nonumber \\
\mathbf{Z} & \rightarrow \left[ \begin{array}{c|c} 1 & 0\end{array}\right].
\label{eq:pauli_mapping}
\end{align}

The evolution of the Pauli operator $P \in \mathcal{P}_n$ over quantum Clifford gates can be described using the conjugation operation. Explicitly, the conjugation of a unitary operator $\mathbf{N}$ under the unitary transformation $\mathbf{M}$ is the unitary transformation $\mathbf{V}$, which is defined as~\cite{gottesman1998heisenberg}
\begin{equation}
\mathbf{V} = \mathbf{{M} \cdot {N} \cdot {M}}^{\dagger}.
\label{eq:conjugation}
\end{equation}
For instance, based on Eq.~\eqref{eq:conjugation}, the conjugation of the Pauli matrix $\mathbf{Z}$ and $\mathbf{X}$ over Hadamard $(\mathbf{H})$ gate is given by
\begin{align}     
\mathbf{H \cdot Z \cdot H}^{\dagger} &= \mathbf{X}, \nonumber \\
\mathbf{H \cdot X \cdot H}^{\dagger} &= \mathbf{Z}. 
\label{eq:hadamard_transform}
\end{align}
Additionally, we can also describe the conjugation of a Pauli matrix $\mathbf{Z}$ and $\mathbf{X}$ over a two-qubit quantum gate, as exemplified by a CNOT gate, as follows: 
\begin{align}
(\mathbf{CNOT}) \cdot (\mathbf{Z} \otimes \mathbf{I}) \cdot (\mathbf{CNOT})^{\dagger} & = \mathbf{Z} \otimes \mathbf{I}, \nonumber \\
(\mathbf{CNOT}) \cdot (\mathbf{I} \otimes \mathbf{Z}) \cdot (\mathbf{CNOT})^{\dagger} & = \mathbf{Z} \otimes \mathbf{Z}, \nonumber \\
(\mathbf{CNOT}) \cdot (\mathbf{X} \otimes \mathbf{I}) \cdot (\mathbf{CNOT})^{\dagger} & = \mathbf{X} \otimes \mathbf{X}, \nonumber \\
(\mathbf{CNOT}) \cdot (\mathbf{I} \otimes \mathbf{X}) \cdot (\mathbf{CNOT})^{\dagger} & = \mathbf{I} \otimes \mathbf{X},
\label{eq:cnot_transform}
\end{align}
where the first Pauli matrix is applied to the control qubit, while the second Pauli matrix is applied to the target qubit.

\begin{figure}
\includegraphics[width=\linewidth]{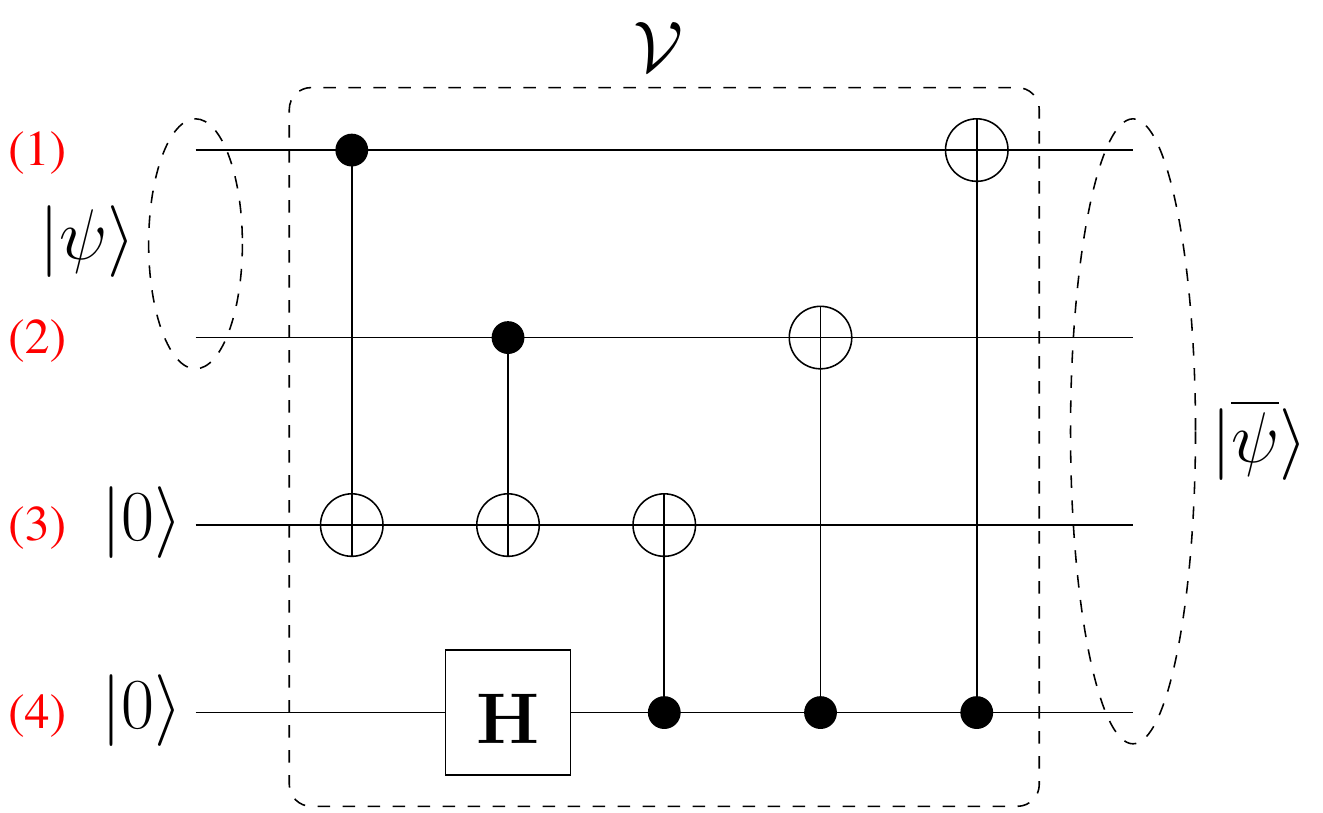}
\caption{The quantum encoder $\mathcal{V}$ of the QSBC $\mathcal{C}[4,2,2]$.}
\label{fig:example}
\end{figure}

Therefore, using the conjugation definition of Eq.~\eqref{eq:conjugation}, we can keep track of the evolution of any $n$-tuple Pauli operator $P \in \mathcal{P}_n$ owing to unitary operations carried out by quantum Clifford gates, such as the QSBC encoders $\mathcal{V}$ and also its inverse encoder $\mathcal{V}^{\dagger}$ illustrated in Fig.~\ref{fig:circuit_layout}. Since the quantum encoders $\mathcal{V}$ of QSBCs are only composed of Hadamard and CNOT gates, we can create a $(2n \times 2n)$-element binary matrix $V$ for classically simulating the evolution of the Pauli operator $P \in \mathcal{P}_n$ over the quantum encoder $\mathcal{V}$. As an example, let us consider the quantum encoder $\mathcal{V}$ of the QSBC $\mathcal{C}[4,2,2]$ seen in Fig.~\ref{fig:example}. 

We commence by initializing $V^{(0)} = \mathbf{I}_{2n}$, where $\mathbf{I}_{2n}$ is a $2n$-dimensional identity matrix formulated as follows:
\begin{equation}
\small
{V}^{(0)} = \left[ \begin{array}{cccc|cccc} 
1 & 0 & 0 & 0 & 0 & 0 & 0 & 0 \\ 
0 & 1 & 0 & 0 & 0 & 0 & 0 & 0 \\
0 & 0 & 1 & 0 & 0 & 0 & 0 & 0 \\
0 & 0 & 0 & 1 & 0 & 0 & 0 & 0 \\
\hline
0 & 0 & 0 & 0 & 1 & 0 & 0 & 0 \\
0 & 0 & 0 & 0 & 0 & 1 & 0 & 0 \\
0 & 0 & 0 & 0 & 0 & 0 & 1 & 0 \\
0 & 0 & 0 & 0 & 0 & 0 & 0 & 1 
\end{array} \right].
\label{eq:v0}
\end{equation}
The $i$-th and the $(n+i)$-th column of matrix $V^{(0)}$ are associated with the evolution of the Pauli matrices $\mathbf{Z}$ and $\mathbf{X}$, respectively, on the $i$-th qubit. 

The first unitary operation in the QSBC encoder $\mathcal{V}$ of Fig.~\ref{fig:example} is the unitary operation $\text{CNOT}(1,3)$, where the notation $\text{CNOT}(i,j)$ means that the $i$-th qubit controls the $j$-th qubit. Now, based on Eq.~\eqref{eq:cnot_transform}, the CNOT unitary transformation propagates the Pauli $\mathbf{X}$ matrix from the control qubit to the target qubit and by contrast, propagates the Pauli $\mathbf{Z}$ matrix from the target qubit to the control qubit. Therefore, in the matrix $V$, the unitary transformation $\text{CNOT}(i,j)$ can be carried out by replacing the $i$-th column with the modulo-2 addition between the $i$-th column and the $j$-th column then replacing the $(n+j)$-th column with the modulo-2 addition between the $(n+i)$-th column and the $(n+j)$-th column. In the case of $V^{(0)}$, the unitary transformation $\text{CNOT}(1,3)$ can simply be viewed as copying the $1$ value from the $3$-rd column to the $1$-st column then copying the value $1$ from $5$-th column to the $7$-th column. Hence, the unitary transformation $\text{CNOT}(1,3)$ transforms the matrix $V^{(0)}$ into the matrix $V^{(1)}$ as follows:
\begin{equation}
\small
{V}^{(1)} = \left[ \begin{array}{cccc|cccc} 
1 & 0 & 0 & 0 & 0 & 0 & 0 & 0 \\ 
0 & 1 & 0 & 0 & 0 & 0 & 0 & 0 \\
\textcolor{red}{\bf 1} & 0 & \textcolor{red}{\bf 1} & 0 & 0 & 0 & 0 & 0 \\
0 & 0 & 0 & 1 & 0 & 0 & 0 & 0 \\
\hline
0 & 0 & 0 & 0 & \textcolor{red}{\bf 1} & 0 & \textcolor{red}{\bf 1} & 0 \\
0 & 0 & 0 & 0 & 0 & 1 & 0 & 0 \\
0 & 0 & 0 & 0 & 0 & 0 & 1 & 0 \\
0 & 0 & 0 & 0 & 0 & 0 & 0 & 1 
\end{array} \right].
\label{eq:v1}
\end{equation}
We indicate the matrix elements involved in the associated transformation using bold red fonts. The unitary transformation $\text{CNOT}(1,3)$ is followed by the second unitary transformation taking place in the QSBC encoder $\mathcal{V}$ of Fig.~\ref{fig:example}, namely the $\text{CNOT}(2,3)$. Following the same method as that used for obtaining the matrix ${V}^{(1)}$, the action of $\text{CNOT}(2,3)$ applied to the matrix ${V}^{(1)}$ yields the matrix ${V}^{(2)}$ as follows:
\begin{equation}
\small
{V}^{(2)} = \left[ \begin{array}{cccc|cccc} 
1 & 0 & 0 & 0 & 0 & 0 & 0 & 0 \\ 
0 & 1 & 0 & 0 & 0 & 0 & 0 & 0 \\
1 & \textcolor{red}{\bf 1} & \textcolor{red}{\bf 1} & 0 & 0 & 0 & 0 & 0 \\
0 & 0 & 0 & 1 & 0 & 0 & 0 & 0 \\
\hline
0 & 0 & 0 & 0 & 1 & 0 & 1 & 0 \\
0 & 0 & 0 & 0 & 0 & \textcolor{red}{\bf 1} & \textcolor{red}{\bf 1} & 0 \\
0 & 0 & 0 & 0 & 0 & 0 & 1 & 0 \\
0 & 0 & 0 & 0 & 0 & 0 & 0 & 1 
\end{array} \right].
\label{eq:v2} 
\end{equation}
The third unitary transformation taking place within the QSBC encoder $\mathcal{V}$ of Fig~\ref{fig:example} is the Hadamard transformation $\mathbf{H}(4)$. Based on Eq.~\eqref{eq:hadamard_transform}, the Hadamard transformation modifies the Pauli matrix $\mathbf{Z}$ into $\mathbf{X}$ and vice versa. Therefore, in a matrix $V$, a Hadamard transformation of $\mathbf{H}(i)$ can be interpreted as swapping the value of the $i$-th column and the $(n+i)$-th column. In case of the matrix $V^{(2)}$, the action of the unitary transformation $\mathbf{H}(4)$ swaps the value of $4$-th column and the $8$-th column, hence resulting in the matrix $V^{(3)}$ as follows:
\begin{equation}
\small
{V}^{(3)} = \left[ \begin{array}{cccc|cccc} 
1 & 0 & 0 & 0 & 0 & 0 & 0 & 0 \\ 
0 & 1 & 0 & 0 & 0 & 0 & 0 & 0 \\
1 & 1 & 1 & 0 & 0 & 0 & 0 & 0 \\
0 & 0 & 0 & 0 & 0 & 0 & 0 & \textcolor{red}{\bf 1} \\
\hline
0 & 0 & 0 & 0 & 1 & 0 & 1 & 0 \\
0 & 0 & 0 & 0 & 0 & 1 & 1 & 0 \\
0 & 0 & 0 & 0 & 0 & 0 & 1 & 0 \\
0 & 0 & 0 & \textcolor{red}{\bf 1} & 0 & 0 & 0 & 0 
\end{array} \right].
\label{eq:v3} 
\end{equation}
These operations are then followed by the unitary transformation $\text{CNOT}(4,3)$, which yields the matrix ${V}^{(4)}$ as follows:
\begin{equation}
\small
{V}^{(4)} = \left[ \begin{array}{cccc|cccc} 
1 & 0 & 0 & 0 & 0 & 0 & 0 & 0 \\ 
0 & 1 & 0 & 0 & 0 & 0 & 0 & 0 \\
1 & 1 & \textcolor{red}{\bf 1} & \textcolor{red}{\bf 1} & 0 & 0 & 0 & 0 \\
0 & 0 & 0 & 0 & 0 & 0 & \textcolor{red}{\bf 1} & \textcolor{red}{\bf 1} \\
\hline
0 & 0 & 0 & 0 & 1 & 0 & 1 & 0 \\
0 & 0 & 0 & 0 & 0 & 1 & 1 & 0 \\
0 & 0 & 0 & 0 & 0 & 0 & 1 & 0 \\
0 & 0 & 0 & 1 & 0 & 0 & 0 & 0 
\end{array} \right].
\label{eq:v4} 
\end{equation}
The action of $\text{CNOT}(4,2)$ upon the matrix $V^{(4)}$ gives us the matrix $V^{(5)}$ as follows:
\begin{equation}
\small
{V}^{(5)} = \left[ \begin{array}{cccc|cccc} 
1 & 0 & 0 & 0 & 0 & 0 & 0 & 0 \\ 
0 & \textcolor{red}{\bf 1} & 0 & \textcolor{red}{\bf 1} & 0 & 0 & 0 & 0 \\
1 & \textcolor{red}{\bf 1} & 1 & \textcolor{red}{\bf 0} & 0 & 0 & 0 & 0 \\
0 & 0 & 0 & 0 & 0 & \textcolor{red}{\bf 1} & 1 & \textcolor{red}{\bf 1} \\
\hline
0 & 0 & 0 & 0 & 1 & 0 & 1 & 0 \\
0 & 0 & 0 & 0 & 0 & 1 & 1 & 0 \\
0 & 0 & 0 & 0 & 0 & 0 & 1 & 0 \\
0 & 0 & 0 & 1 & 0 & 0 & 0 & 0 
\end{array} \right].
\label{eq:v5}
\end{equation}
Finally, applying the $\text{CNOT}(4,1)$ transforms the matrix $V^{(5)}$ into the final matrix $V$ as follows:
\begin{equation}
\small
{V} = \left[ \begin{array}{cccc|cccc} 
\textcolor{red}{\bf 1} & 0 & 0 & \textcolor{red}{\bf 1} & 0 & 0 & 0 & 0 \\ 
0 & 1 & 0 & 1 & 0 & 0 & 0 & 0 \\
\textcolor{red}{\bf 1} & 1 & 1 & \textcolor{red}{\bf 1} & 0 & 0 & 0 & 0 \\
0 & 0 & 0 & 0 & \textcolor{red}{\bf 1} & 1 & 1 & \textcolor{red}{\bf 1} \\
\hline
0 & 0 & 0 & 0 & 1 & 0 & 1 & 0 \\
0 & 0 & 0 & 0 & 0 & 1 & 1 & 0 \\
0 & 0 & 0 & 0 & 0 & 0 & 1 & 0 \\
0 & 0 & 0 & 1 & 0 & 0 & 0 & 0 
\end{array} \right].
\label{eq:v}
\end{equation}
The resultant matrix $V$ is the classical analogue of the quantum encoder $\mathcal{V}$ of the QSBC $\mathcal{C}[4,2,2]$ seen in Fig.~\ref{fig:example}. In order to obtain the matrix $V^{-1}$ of the quantum inverse encoder $\mathcal{V}^{\dagger}$, the same method is invoked. The only difference is that we apply the transformation for each step by reading the quantum circuit from right to the left.

In order to show that the matrix $V$ can be used for classical simulation, let us consider a Pauli operator $P \in \mathcal{P}_4$ as follows:
\begin{equation}
P = \mathbf{Z} \otimes \mathbf{I} \otimes \mathbf{X} \otimes \mathbf{Y}.
\label{eq:pauli_example}
\end{equation}
By using the Pauli-to-binary mapping of Eq.~\eqref{eq:pauli_mapping}, the Pauli operator $P$ given in Eq.~\eqref{eq:pauli_example} can be transformed into its classical analogue as follows:
\begin{equation}
P = \left[ \begin{array}{cccc|cccc} 1 & 0 & 0 & 1 & 0 & 0 & 1 & 1 \end{array} \right].
\end{equation}
The resultant Pauli operator $\widehat{P}$ due to the QSBC encoder $V$ can be obtained by modulo-2 multiplication $(\ast)$ of the vector $P$ and the matrix $V$, which gives us
\begin{align}
\widehat{P} &= P \ast V \nonumber \\
 &= \left[ \begin{array}{cccc|cccc} 1 & 0 & 0 & 0 & 1 & 1 & 0 & 1 \end{array} \right].
\label{eq:pauli_result}
\end{align}
From the resultant vector $\widehat{P}$ of Eq.~\eqref{eq:pauli_result} and also from the Pauli-to-binary mapping of Eq.~\eqref{eq:pauli_mapping}, we can map back the binary vector $\widehat{P}$ into its corresponding Pauli operator, which gives us $\widehat{P} \in \mathcal{P}_4$ as follows:
\begin{equation}
\widehat{P} = \mathbf{Y} \otimes \mathbf{X} \otimes \mathbf{I} \otimes \mathbf{X}.
\end{equation}

In order to simplify the expression of matrix $V$, often the seed transformation $\mathcal{U}$ representation is used. The $U_i$ element of the seed transformation $\mathcal{U} = \lbrace U_1, U_2, \ldots, U_{2n} \rbrace$ is the decimal representation of the $i$-th row of the binary matrix $V$. Therefore, based on the matrix $V$ of Eq.~\eqref{eq:v}, the associated seed transformation $\mathcal{U}$ is given by 
\begin{equation}
\mathcal{U} = \lbrace 144, 80, 240, 15, 10, 6, 2, 16 \rbrace_{10}.
\label{eq:u}
\end{equation}
Finally, the seed transformation $\mathcal{U}$ for all the QSBC encoders of Fig.~\ref{fig:circuit_layout} is provided in Table.~\ref{table:seed}.

\section{Quantum Turbo Code Design Using QSBCs}
\label{Quantum Turbo Code Design Using QSBCs}

QSBCs on their own are only capable of detecting the presence of errors in the state of physical qubits, but not correcting them. This limited capability is due to the minimum distance of $d = 2$ inherited from the classical SBCs. In this treatise, we invoke the QTC scheme utilized in~\cite{poulin2009quantum, babar2015road} as the foundation of developing QSBC-based QTCs. The QSBCs are invoked as the outer codes, while a non-catastrophic and non-recursive QURC is used as the inner code, which we will refer to as the QSBC-QURC scheme for the rest of this treatise. 

The QTC scheme was first introduced by Poulin \textit{et. al} in~\cite{poulin2009quantum}. The proposed QTC utilized two QCCs, which were concatenated in a serial manner. Each of the QCC components exhibits a $1/3$ quantum coding rate and hence, the final quantum coding rate $r_Q$ is $1/9$. To the best of our knowledge, the QTCs operating closest to the quantum hashing bound rely on the construction presented in~\cite{babar2015exit, babar2015road} in the open literature. The near-hashing-bound performance was attained by utilizing QIrCCs both as the outer and the inner codes. Furthermore, the weighting factors of the QIrCC component codes were optimized by invoking EXIT-chart-based heuristic search~\cite{el2014exit, babar2015exit}. Readers who are interested to delve deeper into QTCs and into near-hashing-bound constructions, please refer to~\cite{babar2015road}.

\begin{figure*}[ht!]
\center
\includegraphics[width=0.7\linewidth]{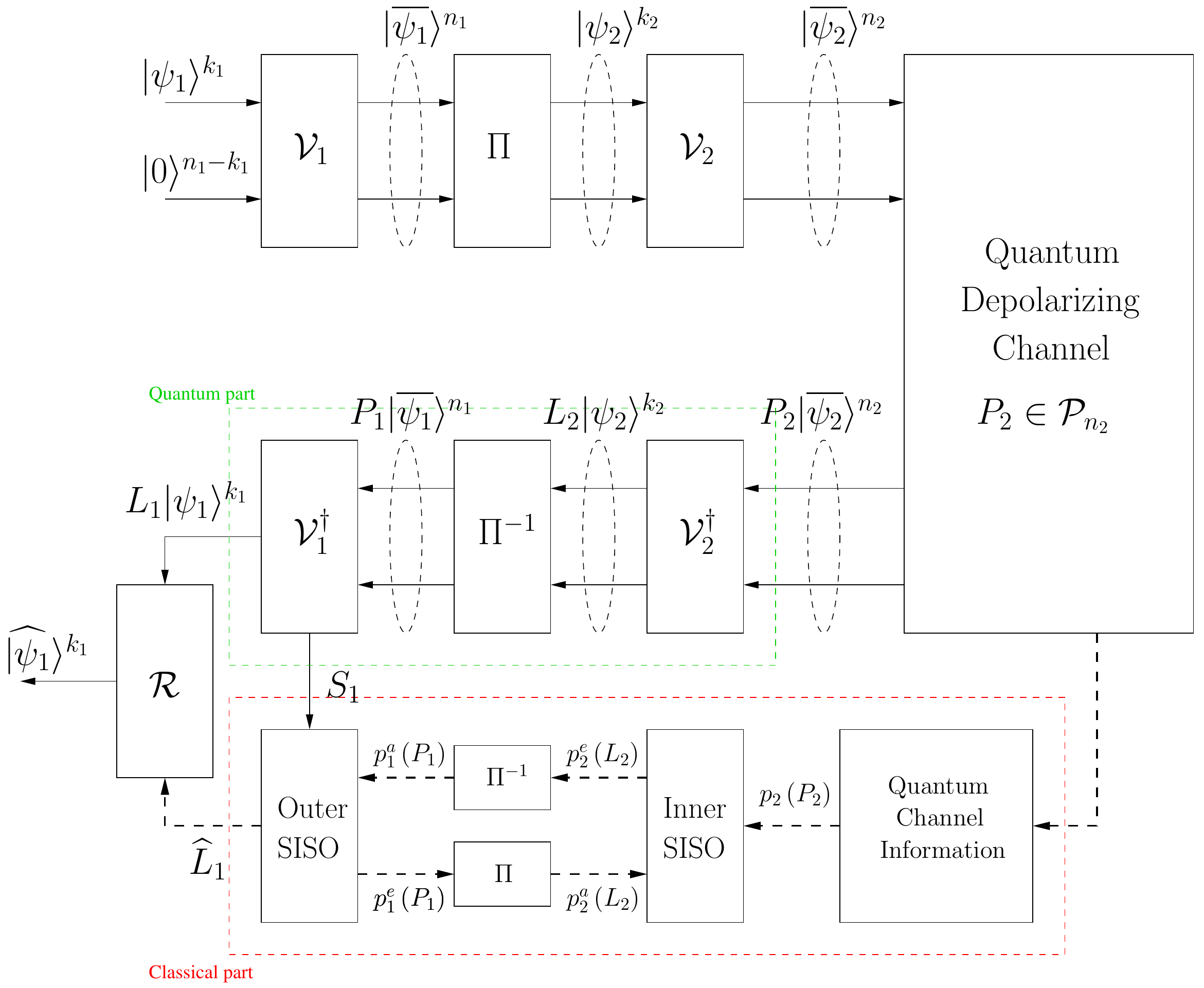}
\caption{The general schematic of serial QTCs utilizing QURC as the inner code. The decoding process, in general, can be separated into two parallel processes, namely the quantum processing part and the classical processing part. The quantum part is represented by the green dashed lines, while the classical part is represented by red the dashed lines.}
\label{fig:qsbc-qurc}
\end{figure*}

\subsection{Encoding Process}
\label{Encoding Process}

In this section, we will describe the proposed QSBC-QURC scheme, whose general schematic can be seen in Fig.~\ref{fig:qsbc-qurc}. The outer encoder $\mathcal{V}_1$ in Fig.~\ref{fig:qsbc-qurc} is a QSBC encoder, which is already shown in Fig.~\ref{fig:circuit_layout}. It maps $k_1$ logical qubits into $n_1$ physical qubits with the aid of $(n_1 - k_1)$ auxiliary qubits according to the following transformation:
\begin{equation}
\mathcal{V}_1 \left( |\psi\rangle^{k_1} \otimes |0\rangle^{\otimes (n_1 - k_1)} \right) = |\overline{\psi_1}\rangle^{n_1}.
\end{equation}
The output of $\mathcal{V}_1$ is fed to the interleaver $\Pi$, which can be represented mathematically as a permutation matrix and can be realized physically as a series of quantum SWAP gates. The interleaving process can be formally written as 
\begin{equation}
\Pi \left( |\overline{\psi_1}\rangle^{n_1} \right) = |\psi_2\rangle^{k_2},
\end{equation}
where we have $k_2 = n_1$, since the interleaver does not alter the number of physical qubits, only rearranges the position of the qubits indices in the quantum state. Hence, the Hamming weight of the state of physical qubits is not changed after this process. Next, the output of the interleaver $\Pi$ is fed into the inner encoder $\mathcal{V}_2$, which carries out the following transformation: 
\begin{equation}
\mathcal{V}_2 \left( |\psi\rangle^{k_2} \otimes |0\rangle^{\otimes (n_2 - k_2)} \right) = |\overline{\psi_2}\rangle^{n_2}.
\label{eq:encoder}
\end{equation}
The encoder $\mathcal{V}_2$ maps the state of $k_2$ logical qubits into the state of $n_2$ physical qubits with the aid of $(n_2 - k_2)$ auxiliary qubits. Since, we are employing the QURCs as the inner codes, the transformation in Eq.~\eqref{eq:encoder} can be further simplified as
\begin{equation}
\mathcal{V}_2 \left( |\psi\rangle^{k_2} \right) = |\overline{\psi_2}\rangle^{n_2}.
\label{eq:encoder_urc}
\end{equation}
The main difference between the interleaver and the QURC is that the state of physical qubits after the QURC may experience Hamming weight alterations.

\subsection{Quantum Depolarizing Channel}
\label{Quantum Depolarizing Channel}

After the encoding process, the encoded state of physical qubits may experience quantum decoherence. In this study, we use the quantum depolarizing channel~\cite{nielsen2000quantum}. This depolarizing channel models the imperfection of quantum gates, as well as the coherence taking place in the quantum memory, and even the actual quantum transmission channel through free space or optical fiber channels. The quantum decoherence is represented by the $n$-tuple Pauli operator $P_2 \in \mathcal{P}_n$ and its action imposed upon the encoded state of physical qubits $|\overline{\psi_2}\rangle^{n_2}$ can be expressed as
\begin{equation}
|\widetilde{\psi}\rangle = P_2 \left( |\overline{\psi_2}\rangle^{n_2} \right).
\end{equation}
The error operator $P_2$ is characterized by the depolarizing probability $p$. To elaborate a little further, the error operator $P \in \mathcal{P}_{n}$ is an $n$-tuple Pauli operator, where each qubit may independently experience a bit-flip ($\mathbf{X}$) error, a phase-flip ($\mathbf{Z}$) error as well as a simultaneous bit-flip and phase flip ($\mathbf{Y}$) error. The probability of each qubit experiencing an $\mathbf{X}$, $\mathbf{Z}$, and $\mathbf{Y}$ error is denoted by $p_{\mathbf{X}}$, $p_{\mathbf{Z}}$, and $p_{\mathbf{Y}}$, respectively. Under the assumption that $p_{\mathbf{X}} +p_{\mathbf{Y}} + p_{\mathbf{Z}} = p$ and $p_{\mathbf{X}} = p_{\mathbf{Y}} = p_{\mathbf{Z}} = p/3$, this quantum channel is referred to as \textit{symmetric quantum depolarizing channel}~\cite{nielsen2000quantum}. Needless to say, it is always possible to create a model where we have the assumption that $p_{\mathbf{X}} \neq p_{\mathbf{Y}} \neq p_{\mathbf{Z}}$, which can be deeemed to be more realistic~\cite{nguyen2016exit}. However, choosing the value such as $p_{\mathbf{X}} = p_{\mathbf{Y}} = p_{\mathbf{Z}}$ will provide us with the worst-case scenario, because we have to provide the same level of protection for different types of errors without favoring only one specific type of error, which can result in quantum coding rate or QBER improvements. A more detailed discourse on QSC design for asymmetric quantum depolarizing channels, enthusiastic readers might like to refer to~\cite{ioffe2007asymmetric, sarvepalli2009asymmetric, wang2010asymmetric, nguyen2016exit}.

\subsection{Decoding Process}
\label{Decoding Process}

Generally speaking, the decoding process of any QSC relies on the conjunction of two parts, namely the quantum information processing part and the classical information processing part. In Fig.~\ref{fig:qsbc-qurc}, the quantum processing part is marked by the components bounded by the green dashed lines, while the classical processing part is represented by the components bounded by the red dashed lines. First, let us describe the quantum processing part. The corrupted state of physical qubits $|\overline{\psi_2}\rangle^{n_2}$ is fed to the inverse encoder $\mathcal{V}_2^{\dagger}$ of Fig.~\ref{fig:qsbc-qurc}, which represents the conjugate transpose of encoder $\mathcal{V}_2$. Physically, they can be implemented identically with the only difference is that the input and output of the inverse encoder $\mathcal{V}^{\dagger}$ is in the reverse position compared to the encoder $\mathcal{V}$. Since the quantum encoder $\mathcal{V}$ and its inverse encoder $\mathcal{V}^{\dagger}$ are composed by the quantum Clifford gates and the error operator $P$ is an $n$-tuple Pauli operator $P \in \mathcal{P}_n$, the act of inverse encoders $\mathcal{V}_2^{\dagger}$ will decompose the error operator $P_2$ into two error components as follows:  
\begin{equation}
\mathcal{V}^{\dagger}_2 \left( P_2 \left( |\overline{\psi_2}\rangle^{n_2} \right) \right) = L_2 |\overline{\psi_2}\rangle^{k_2} \otimes S_2|0\rangle^{n_2 - k_2},
\label{eq:inverse}
\end{equation}
where $L_2$ is the error operator on $k_2$ logical qubits and $S_2$ is the error operator on $(n_2 - k_2)$ auxiliary qubits. The auxiliary qubits are then measured in the relevant computational basis, where the value $S_2$ can be treated as a syndrome in classical error correction codes, which is forwarded to the classical processing part. However, since the inverse encoder of $\mathcal{V}_2^{\dagger}$ is an inverse encoder of a QURC, Eq.~\eqref{eq:inverse} can be further simplified to
\begin{equation}
\mathcal{V}^{\dagger}_2 \left( P_2 \left( |\psi_2\rangle^{n_2} \right) \right) = L_2 |\psi_2\rangle^{k_2},
\label{eq:inverse_urc}
\end{equation}
since we have $k_2 = n_2$. Next, the output of the inverse encoder $\mathcal{V}_2^{\dagger}$ is passed trough the deinterleaver $\Pi^{-1}$. This transformation can be formally expressed as
\begin{equation}
\Pi^{-1} \left( L_2 |\overline{\psi_2}\rangle^{k_2} \right) = P_1|\overline{\psi_1}\rangle^{n_1}.
\end{equation}
The output of the deinterleaver is then processed as the input of $\mathcal{V}_1^{\dagger}$, which is subjected to an identical transformation as $\mathcal{V}_2^{\dagger}$. This can be expressed as follows:
\begin{equation}
\mathcal{V}^{\dagger}_1 \left( P_1 \left( |\overline{\psi_1}\rangle^{n_1} \right) \right) = L_1 |\psi_1\rangle^{k_1} \otimes S_1|0\rangle^{n_1 - k_1}.
\end{equation}
In this QSBC-QURC scheme, the inverse encoder of $\mathcal{V}_1^{\dagger}$ is constituted by the quantum inverse encoder $\mathcal{V}^{\dagger}$ of the QSBC, which is implemented by flipping the input and ouput of the quantum encoder $\mathcal{V}$ seen in Fig.~\ref{fig:circuit_layout}.

Finally, based on the information obtained from the classical information processing part, the error recovery operator $\mathcal{R}$ is applied to the output of the inverse encoder  $\mathcal{V}_1^{\dagger}$ in order to obtain the predicted logical qubit state as follows:
\begin{equation}
\mathcal{R}\left( L_1 |\psi_1\rangle^{k_1} \right) = |\widehat{\psi_1}\rangle^{k_1}.
\label{eq:recovery}
\end{equation}
If $\mathcal{R} = L_1$, we obtain $|\widehat{\psi_1}\rangle^{k_1} = |\psi_1\rangle^{k_1}$, which completes our decoding process.

Let us now take a step back to elaborate a little further on the classical processing part of the decoding process. The classical decoder part for a QTC is very similar to that of classical turbo codes. It consists of two soft-input soft-output (SISO) decoders, an interleaver, and a deinterleaver. 

As seen in Fig.~\ref{fig:qsbc-qurc}, the classical processing is started by 
obtaining the quantum depolarizing probability $p$ of the quantum channel associated with the error operator $P_2$. In this work, we assume that we have perfect knowledge of the quantum depolarizing probability $p$. The depolarizing probability value $p$ and the \textit{a priori} information $p^a_2\left( L_2 \right)$ obtained from the outer SISO decoder are used by the inner SISO decoder for calculating the extrinsic information $p^e_2\left( L_2 \right)$. For the first iteration, the depolarizing probability $p$ is the only input value used by the inner SISO decoder. Hence, the value of $p^a_2\left( L_2 \right)$ is initialized to be equiprobable. Next, the extrinsic information $p^e_2\left( L_2 \right)$ is interleaved in order to obtain the \textit{a priori} information $p^a_1\left( P_1 \right)$ for the outer SISO decoder. By combining the \textit{a priori} information $p^a_1\left( P_1 \right)$ and the syndrome value $S_1$, the outer SISO decoder calculates the extrinsic information $p^e_1\left( P_1 \right)$. The extrinsic value is then deinterleaved to yield $p^a_2\left( L_2 \right)$ which is fed into the inner SISO decoder. This process is performed iteratively until one of the following conditions is satisfied: the converged mutual information is attained or the maximum affordable number of iterations is reached. On the final iteration, the outer SISO decoder will produce $\widehat{L}_1$, which is the most likely error pattern, given the value of $p$ and $S_1$ provided by the quantum processing part. The value of $\widehat{L}_1$ is obtained for performing error recovery, as detailed in Eq.~\eqref{eq:recovery}. A more rigorous treatment on the classical processing part of QTCs can be found in~\cite{wilde2014entanglement, babar2015road}

\section{Results and Analysis}
\label{Results and Analysis}

In this section, firstly, we analyze the performance of the QSBC-QURC conceived using EXIT charts~\cite{ten2001convergence, kliewer2006efficient, el2014exit}. In the classical domain, EXIT charts constitute a powerful tool, which is often used for guiding the design of near-capacity iterative error correction and for predicting their performance. An initial encouraging effort conducted in~\cite{babar2015exit} invoked EXIT charts for predicting the performance of iterative QTCs demonstrating that EXIT charts can be indeed extended to the quantum domain. Secondly, we proceed by characterizing the performance of the QSBC-QURC scheme in terms of its quantum bit error ratio (QBER), which we obtained from our Monte Carlo simulations. Thirdly, we translate the QBER performance to the distance from the quantum hashing bound, which directly corresponds to the efficiency of quantum channel utilization, which is also related to the goodput. Finally, we use the goodput metric for determining the depolarizing probability at which switching to different quantum coding rate becomes beneficial for conceiving a multi-rate QSBC-QURC scheme. 

\subsection{EXIT Chart}
\label{EXIT Chart}

In the classical domain, the encoders exhibiting recursive and non-catastrophic properties are highly desirable for conceiving near-capacity turbo codes. Unfortunately, in the quantum domain, the QCCs cannot be simultaneously recursive and non-catastrophic~\cite{houshmand2013recursive}. The recursive structure of QCCs is required for ensuring the convergence of iterative decoding to a vanishingly low QBER. Additionally, the QCCs exhibiting catastrophic structure require a doping mechanism or entanglement-assisted solution in order to substantially benefit from iterative decoding, since catastrophic QCCs provide zero \textit{a priori} information~\cite{banerjee2005nonsystematic, wilde2014entanglement}. These two solutions are beyond the scope of our discussions in this paper. Fortunately, a non-recursive and non-catastrophic QCCs can still be designed for striking an attractive compromise, since they can achieve beneficial iteration gains even if the inner decoder EXIT curve terminates at the $(1,y)$ point for $y < 1$, provided that it only intersects with the outer decoder EXIT curves near $x = 1$~\cite{babar2016serially}. Based on these conditions, an exhaustive EXIT-chart-based heuristic search has been conducted to find a ``good" QURC. The resultant seed transformation for such a QURC is given by
\begin{equation}
\mathcal{U} = \lbrace 21, 56, 5, 46, 44, 38 \rbrace_{10}.
\label{eq:seed}
\end{equation}
In this treatise, our QSBC-QURC scheme utilized a specific a QURC whose seed transformation is given in Eq.~\eqref{eq:seed} and the QSBCs of $\mathcal{C}[4,2,2]$, $\mathcal{C}[6,4,2]$ and $\mathcal{C}[8,6,2]$ were used as the outer codes. The seed transformations of the QSBCs are given in Table~\ref{table:seed}. As a benchmark, we use the the QIrCC-QURC scheme presented in~\cite{babar2016serially}, where the QIrCCs are optimized using EXIT-chart-aided method specified in~\cite{babar2015exit, babar2015road}. The seed transformation of the QIrCC component codes is given in Table~\ref{table:qircc}. As we have described briefly in Subsection~\ref{Classical Simulation for QSBCs}, the seed transformation is the decimal representation used for describing the quantum gate connections amongst the physical qubits within the quantum encoder $\mathcal{V}$. Also, it can be used for simulating the QSCs classically.

\begin{table*}[ht!]
\renewcommand{\arraystretch}{1.5}
\caption{The seed transformation $\mathcal{U}$ associated with QSBCs having various quantum coding rates $r_Q$.\label{table:seed}}
\centering
\small
\begin{tabular}{|c|r|}
\hline
\textbf{Quantum coding rate} $(r_Q)$ & \multicolumn{1}{c|}{Seed transformation $\mathcal{U}$}\\
\hline
\hline
 $1/2$ & $\lbrace 144, 80, 240, 15, 10, 6, 2, 16 \rbrace_{10}$ \\
\hline
 $2/3$ & $\lbrace 2112, 1088, 576, 320, 4032, 63, 34, 18, 10, 6, 2, 64 \rbrace_{10}$ \\
\hline
 $3/4$ & $\lbrace 33024, 16640, 8448, 4352, 2304, 1280, 65280, 255, 130, 66, 34,18, 10, 6, 2, 256 \rbrace_{10}$ \\
\hline
\end{tabular}
\end{table*}

\begin{table*}[ht!]
\renewcommand{\arraystretch}{1.5}
\caption{The seed transformation $\mathcal{U}$ associated with QCCs exhibiting various quantum coding rates $r_Q$ for constructing the QIrCC of~\cite{babar2016serially}. All of the QCCs exhibiting a memory of $m = 3$.\label{table:qircc}}
\centering
\small
\begin{tabular}{|c|r|}
\hline
\textbf{Quantum coding rate} $(r_Q)$ & \multicolumn{1}{c|}{Seed transformation $\mathcal{U}$}\\
\hline
\hline
 $1/4$ & $\lbrace 9600, 691, 11713, 4863, 1013, 6907, 1125, 828, 10372, 6337, 5590, 11024, 12339, 3439 \rbrace_{10}$ \\
\hline
 $1/3$ & $\lbrace 3968, 1463, 2596, 3451, 1134, 3474, 657, 686, 3113, 1866, 2608, 2570 \rbrace_{10}$ \\
\hline
 $1/2$ & $\lbrace 848, 1000, 930, 278, 611, 263, 744, 260, 356, 880 \rbrace_{10}$ \\
\hline
 $2/3$ & $\lbrace 529, 807, 253, 1950, 3979, 2794, 956, 1892, 3359, 2127, 3812, 1580 \rbrace_{10}$ \\
\hline
 $3/4$ & $\lbrace 62, 6173, 4409, 12688, 7654, 10804, 1763, 15590, 6304, 3120, 2349, 1470, 9063, 4020 \rbrace_{10}$ \\
\hline
\end{tabular}
\end{table*}

\begin{figure}[ht!]
\center
\includegraphics[width=\linewidth]{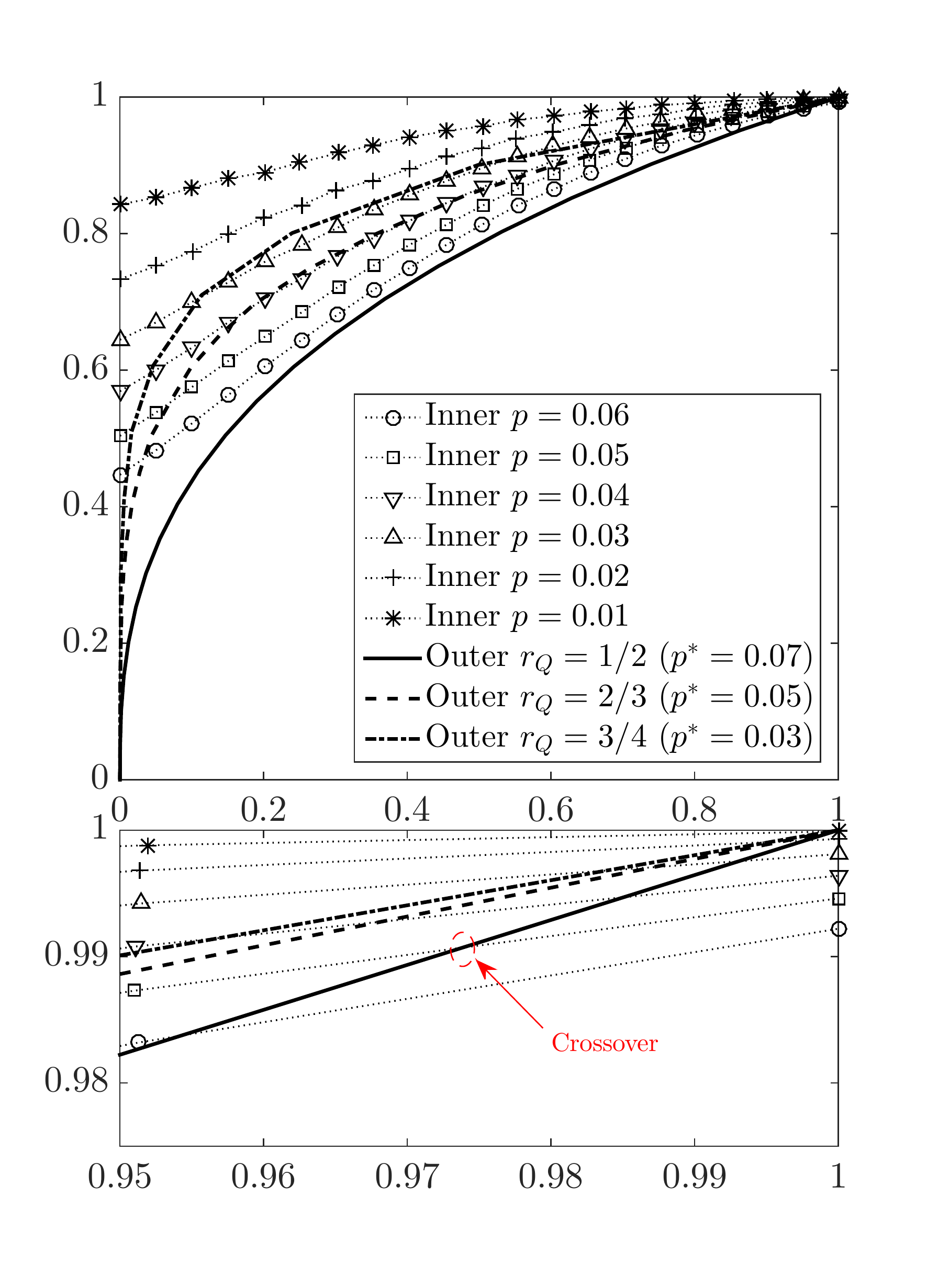}
\caption{Outer decoder EXIT curves for the QSBCs used as the outer codes and the inner decoder EXIT curves for the QURC as the inner code. We can take a closer look at the vicinity of the $(1,1)$ point and observe that despite having been carefully selected, the inner decoder EXIT curve does not reach the $(1,1)$ point due to the nature of the non-recursive structure.}
\label{fig:exitchart}
\end{figure}

\begin{figure}[ht!]
\center
\includegraphics[width=\linewidth]{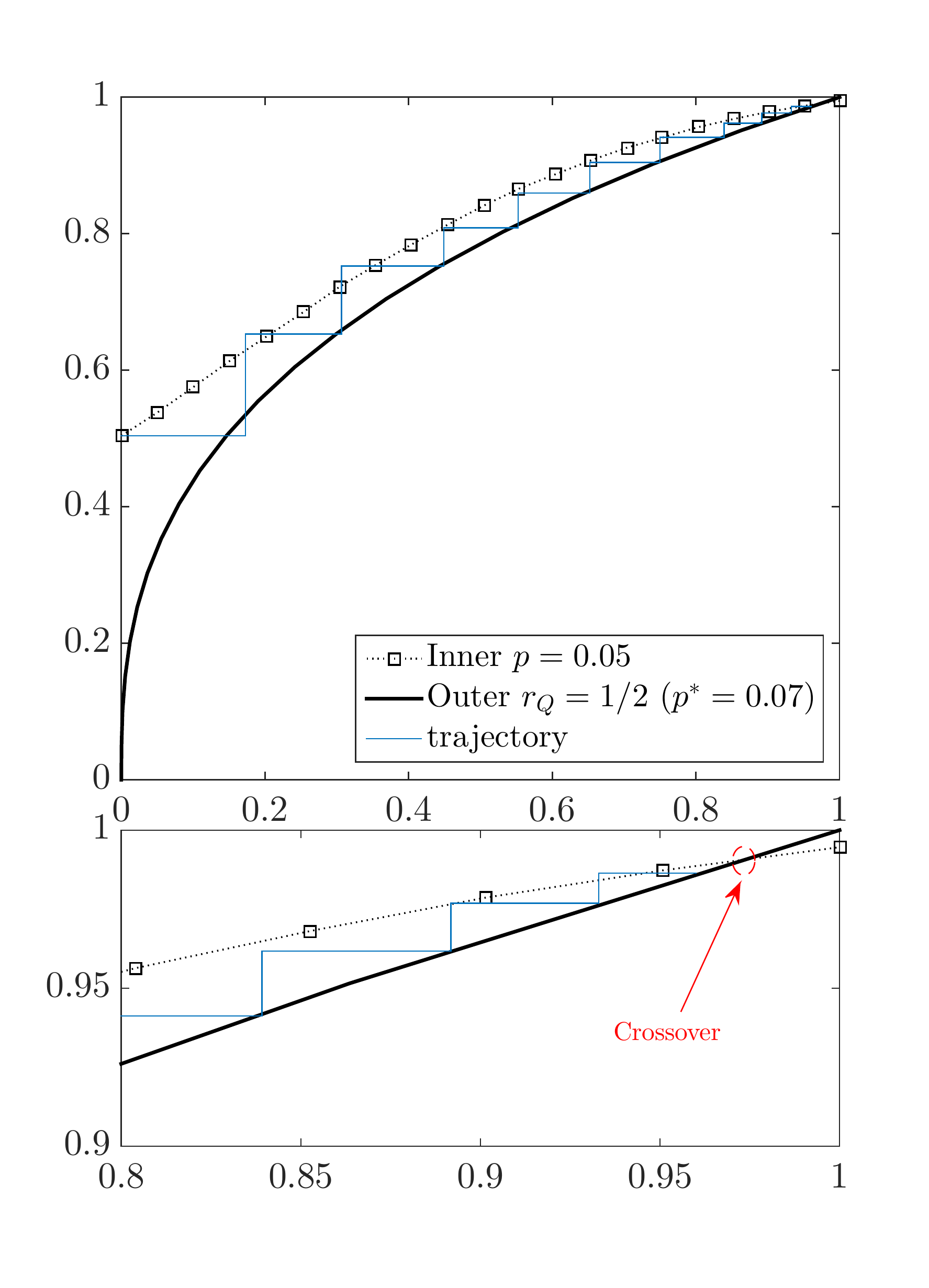}
\caption{The Monte Carlo simulation-based decoding trajectory of mutual information exchange between the classical inner and outer SISO decoders. It can be observed that the decoding trajectory is stuck at the vicinity of $(1,1)$-point after approximately 10 decoding iterations.}
\label{fig:trajectory24}
\end{figure}

The QURC we chose, which is defined by the seed transformation in Eq.~\eqref{eq:seed}, has a non-recursive and non-catastrophic structure. Therefore, the inner decoder EXIT curve will terminate at the $(1,y)$ point, where $y < 1$. In Fig.~\ref{fig:exitchart}, we have plotted the inner decoder EXIT curves for QURCs in the face of various depolarizing probabilities of $p = \lbrace 0.06, 0.05, 0.04, 0.03, 0.02, 0.01 \rbrace$. At a quick glance, the inner decoder EXIT curves of the QURC are seen to be capable of reaching the $(1,1)$ point. However, a closer inspection in the vicinity of the $(1,1)$ point reveals that indeed the inner decoder EXIT curves terminate at the $(1,y)$ point, where $y < 1$. Hence, in order to attain an infinitesimally low QBER, the condition that the intersection of the inner and outer decoder EXIT curves has to be in the proximity of the $(1,1)$ point is no longer trivial. Therefore, in Fig.~\ref{fig:exitchart}, we have also plotted the outer decoder EXIT curves for QSBCs having the quantum coding rates of $r_Q = \lbrace 1/2, 2/3, 3/4 \rbrace$. We can observe that the intersections of the inner and outer decoder EXIT curves for various quantum coding rates $r_Q$ are in the proximity of the $(1,1)$ point, as desired. Furthermore, a marginally open EXIT tunnel emerges between the inner decoder EXIT curve for $p = 0.05$ and the outer decoder EXIT curve for a $1/2$-rate QSBC. This indicates that convergence of the iterative decoding is attained at $p \leq 0.05$, which exhibits itself as waterfall region in the QBER curves. Additionally, in Fig.~\ref{fig:trajectory24}, we have plotted the stair-case-shaped Monte Carlo simulation-based decoding trajectory of mutual information exchange between the inner and outer SISO decoders. As expected, the decoding trajectory got stuck after approximately 10 iterations between the inner and outer SISO decoder in the vicinity of the $(1,1)$ point. Given that the iterative decoding of the QSBC-QURC scheme technically does not achieve a full convergence, we expected an error-floor to be present in the QBER curves, which we will discuss in the next subsection.

\subsection{Quantum Bit Error Rate (QBER)}
\label{Quantum Bit Error Rate (QBER)}

\begin{figure}[ht!]
\center
\includegraphics[width=\linewidth]{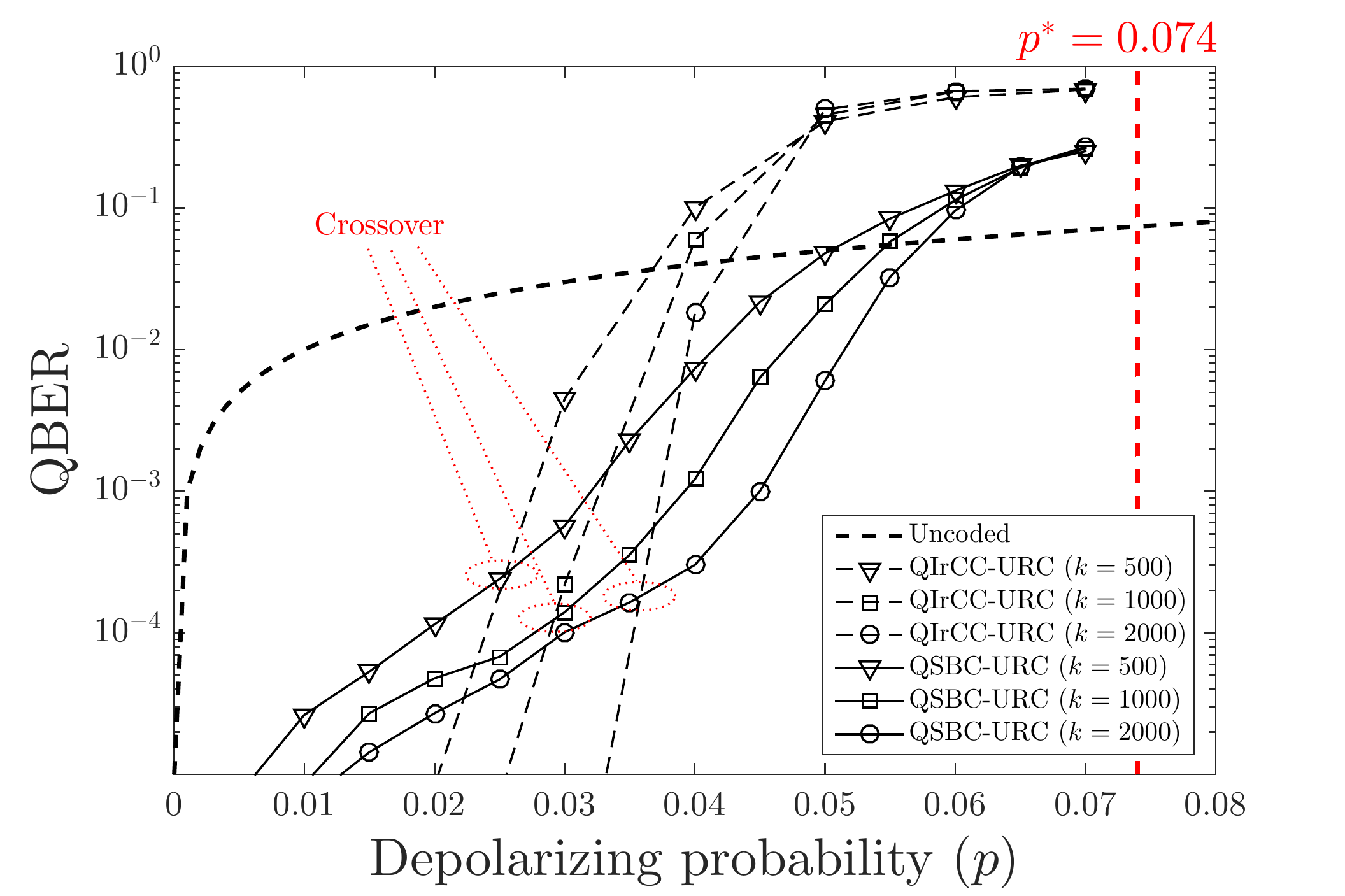}
\caption{QBER versus quantum depolarizing probability curves for the half-rate QSBC-QURC and the half-rate QIrCC-QURC scheme exhibiting various numbers of logical qubits after 16 decoding iterations. For a half-rate QSC, the quantum hashing bound is $p^{\ast} = 0.074$.}
\label{fig:compare1}
\end{figure}

Let us now evaluate the error correction performance of the conceived QSBC-QURC schemes based on their QBER and the distance with respect to the quantum hashing bound. The QBER curves of the half-rate QSBC-QURC scheme having $n = \lbrace 500, 1000, 2000 \rbrace$ physical qubits after 16 decoding iterations using Monte Carlo simulations is portrayed in Fig~\ref{fig:compare1}. For the sake of benchmarking, we also simulated the QIrCC-QURC scheme from~\cite{babar2016serially} and we added the QBER curves to Fig.~\ref{fig:compare1}. Beneficial performance improvements can be observed for both the QSBC-QURC and the QIrCC-QURC arrangements upon increasing the number of physical qubits. 

Let us now compare the QBER performance of the QSBC-QURC and the QIrCC-QURC schemes, where the latter is the most powerful half-rate QTC scheme at the time of writing exhibiting the best QBER performance. It can be observed from Fig.~\ref{fig:compare1} that for $n = 2000$ physical qubits, the proposed QSBC-QURC scheme offer a substantial performance improvement in the depolarizing probability region of $0.035 < p < 0.055$. However, as we reduce the depolarizing probability of the quantum channel to the region of $p < 0.035$, the QIrCC-QURC scheme outperforms the QSBC-QURC, where the latter has a relatively high error floor. There are two possible explanations for this specific phenomenon exhibited by our design. Firstly, based on our EXIT chart analysis, we already expected the emergence of an error floor since the stair-case-shaped decoding trajectory of the QSBC-QURC scheme got stuck before reaching the $(1,1)$-point of perfect convergence. However, based on this argument, the QIrCC-QURC should also have an error floor. Indeed, in reality the QIrCC-QURC scheme is also expected to have an error floor, but at a very low QBER, which is actually unobservable in Fig.~\ref{fig:compare1}. This brings us to the second reason, which also explains why the QSBC-QURC scheme has a significantly higher error floor than the QIrCC-QURC scheme. The answer is related to the characteristics of the outer codes. Explicitly, as for the QIrCC-QURC scheme, the outer code is constituted by a set of QCCs exhibiting strong error correction performance despite failing to converge fully. The seed transformation of the QCCs that assembles the QIrCC is given in Table~\ref{table:qircc}~\cite{babar2016serially}. For more detailed descriptions on QIrCCs, we refer the motivated reader to~\cite{babar2015road}. By contrast, the outer code for our QSBC-QURC schemes are constituted by QSBCs having a minimum distance of $d = 2$. Compared to QIrCCs, QSBCs are the weaker codes. Consequently, it results in residual qubit errors even in the region of low depolarizing probability $p$. Our QBER performance comparison between the QIrCC-QURC and QSBC-QURC schemes is summarized in Table~\ref{table:results1}. Once again, we want to highlight that the QSBC-QURC outperforms the QIrCC-QURC scheme for the scenarios of $\text{QBER} < \text{QBER}_{\text{uncoded}}$ and $\text{QBER} < 10^{-3}$. However, due to the relatively high error floor of the QSBC-QURC scheme, for a scenario where $\text{QBER} < 10^{-4}$ is required, the QIrCC-QURC scheme succeeds in meeting this requirement at a higher depolarizing probability $p$. 

\begin{figure}[ht!]
\center
\includegraphics[width=\linewidth]{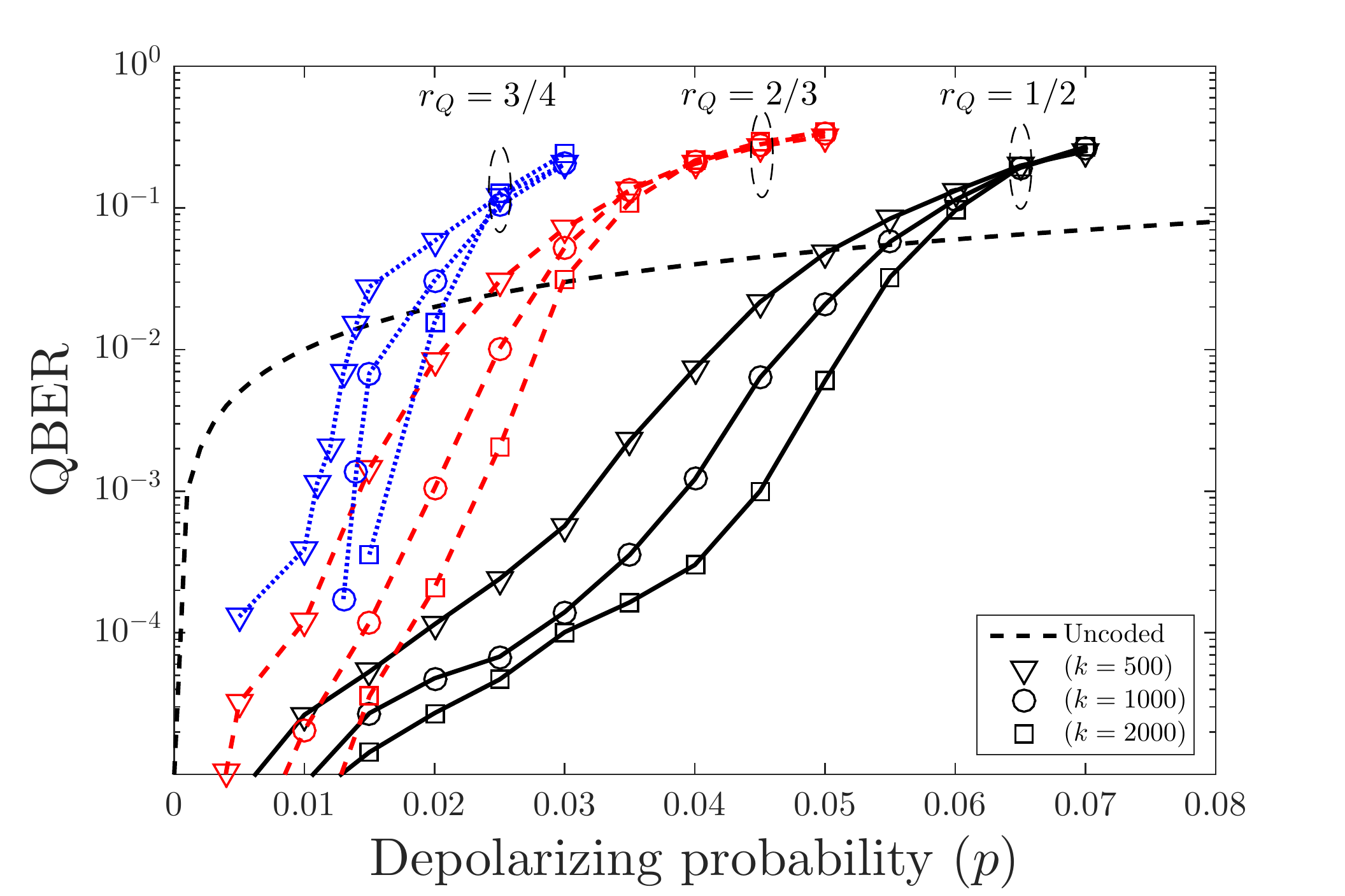}
\caption{QBER versus quantum depolarizing probability curves for QSBC-QURC schemes exhibiting quantum coding rates of $r_Q = \lbrace 1/2, 2/3, 3/4 \rbrace$ having $k = \lbrace 500, 1000, 2000 \rbrace$ logical qubits after 16 decoding iterations.}
\label{fig:compare3}
\end{figure}

However, the main problem with using QIrCCs as the outer codes is that for each quantum coding rate $r_Q$, it requires another exhaustive search for finding the best code and the resultant codes may not share the same encoder structure. By contrast, the QSBCs having various quantum coding rates $r_Q$ share the same quantum encoder structure, as illustrated in Fig.~\ref{fig:circuit_layout}. Therefore, in terms of flexibility and adaptivity, the QIrCCs-QURC schemes may not be favourable. Since the QSBC-QURC schemes can be configured for various $r_Q$ values using the same quantum encoder, we have to further investigate the performance of the QSBC-QURC scheme exhibiting various $r_Q$ values. In Fig.~\ref{fig:compare3}, we have plotted the QBER performance of the QSBC-QURC scheme having quantum coding rates of $r_Q = \lbrace 1/2, 2/3, 3/4 \rbrace$. Naturally, we can go beyond $r_Q = 3/4$, however, our QTC simulations are limited by the computational power of our classical computers. Figure~\ref{fig:compare3} shows that we can reduce the error floor by increasing the number of physical qubits. It also shows that the QSBC-QURC scheme exhibiting a higher quantum coding rate can only cope with a lower quantum depolarizing probability $p$. However, by increasing the quantum coding rate $r_Q$, the effective throughput of the quantum depolarizing channel can be improved, since it requires a lower number of auxiliary qubits, which will be discussed further in the next subsection. The performance results of QSBC-QURC schemes exhibiting various quantum coding rates are summarized in Table~\ref{table:results2}.

\begin{figure}[ht!]
\center
\includegraphics[width=\linewidth]{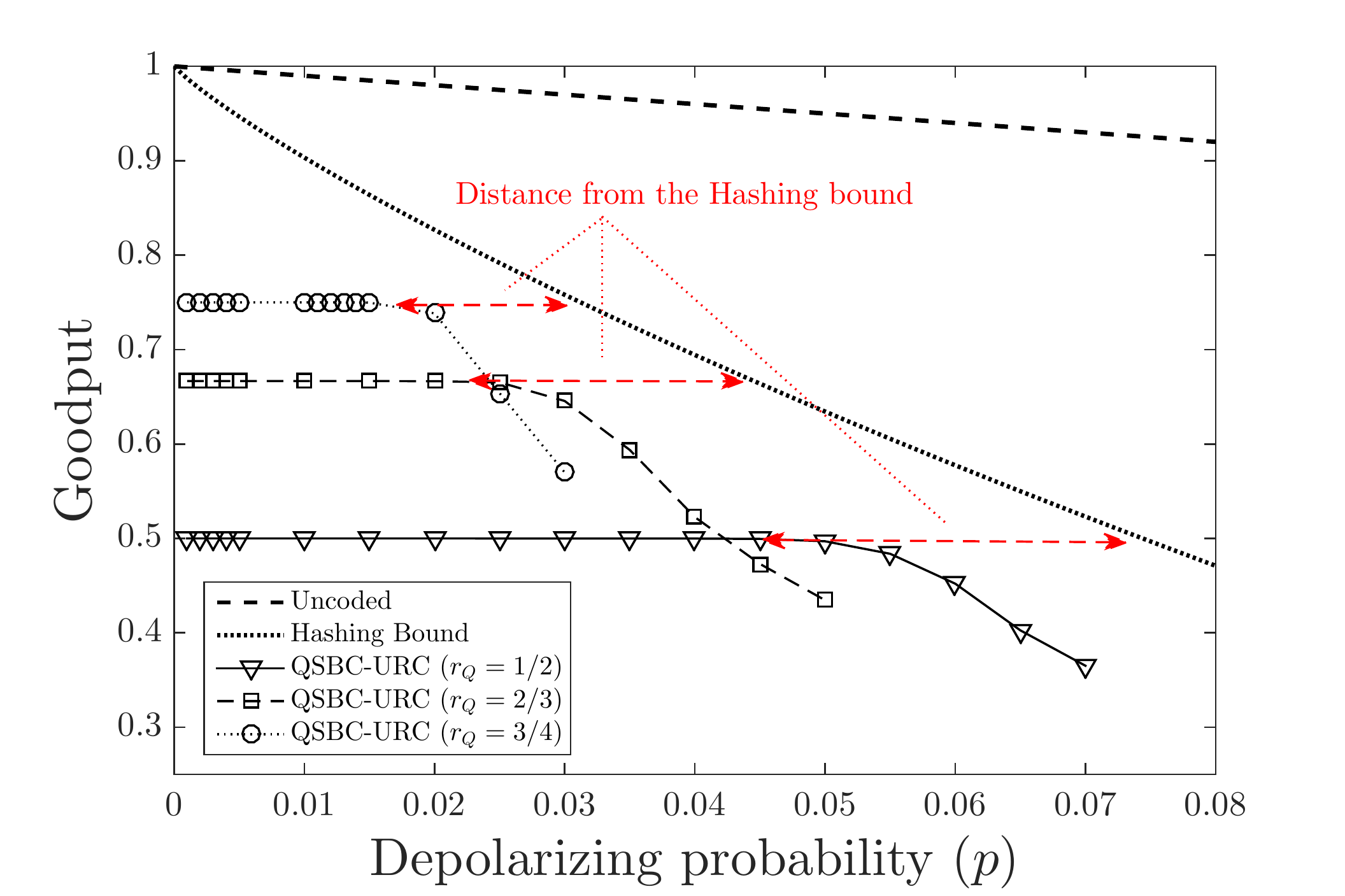}
\caption{Goodput curves versus depolarizing probability for QSBC-URC exhibiting various quantum coding rates having logical qubits $k = 2000$ after 16 decoding iterations. The red dashed lines represent the distance to the quantum hashing bound for each quantum coding rate.}
\label{fig:compare5}
\end{figure}

\begin{table*}[ht!]
\renewcommand{\arraystretch}{1.5}
\caption{Performance comparison of half-rate QIrCC-QURC to half-rate QSBC-QURC for $k = \lbrace 500, 1000, 2000 \rbrace$ logical qubits. The performance is described using the maximum tolerable depolarizing probability $p$ given various requirements, below which the code improves the QBER. For any half-rate QSC, the quantum hashing bound is $p^{\ast} = 0.074$.\label{table:results1}}
\centering
\begin{tabular}{|l||r|r||r|r||r|r|}
\hline
\multirow{2}{*}{\textbf{Requirement}} & \multicolumn{2}{c||}{$k = 500$}& \multicolumn{2}{c||}{$k = 1000$}& \multicolumn{2}{c|}{$k = 2000$} \\
\cline{2-7}
 & \multicolumn{1}{c|}{$p$ QIrCC} & \multicolumn{1}{c||}{$p$ QSBC} & \multicolumn{1}{c|}{$p$ QIrCC} & \multicolumn{1}{c||}{$p$ QSBC} & \multicolumn{1}{c|}{$p$  QIrCC} & \multicolumn{1}{c|}{$p$ QSBC}\\
\hline
\hline
$\text{QBER} < \text{uncoded QBER}$ & 0.037 & \textbf{0.050} & 0.039 & \textbf{0.055} & 0.043 & \textbf{0.058} \\
\hline
$\text{QBER} = 10^{-3}$ & 0.028 & \textbf{0.032} & 0.033 & \textbf{0.039} & 0.037 & \textbf{0.045} \\
\hline
$\text{QBER} = 10^{-4}$ & \textbf{0.024} & 0.019 & \textbf{0.029} & 0.028 & \textbf{0.035} & 0.030 \\
\hline
\end{tabular}
\end{table*}

\begin{table*}[ht!]
\renewcommand{\arraystretch}{1.5}
\caption{Performance comparison of QSBC-QURC schemes having $k = \lbrace 500, 1000, 2000 \rbrace$ logical qubits and quantum coding rates of $r_Q = \lbrace 1/2, 2/3, 3/4 \rbrace$. The performance is described using the maximum tolerable depolarizing probability $p$ given various requirements, below which the code improves the QBER performance. The quantum hashing bound for QSCs having $r_Q = \lbrace 1/2, 2/3, 3/4 \rbrace$ is given by $p^{\ast} = \lbrace 0.074,0.044,0.031 \rbrace$.\label{table:results2}}
\centering
\begin{tabular}{|l||r|r|r||r|r|r|}
\hline
\multirow{2}{*}{\textbf{Requirement}} & \multicolumn{3}{c||}{$\text{QBER} < \text{uncoded QBER}$ }& \multicolumn{3}{c|}{$\text{QBER} = 10^{-3}$}\\
\cline{2-7}
 & \multicolumn{1}{c|}{$r_Q = 1/2$} & \multicolumn{1}{c|}{$r_Q = 2/3$} & \multicolumn{1}{c||}{$r_Q = 3/4$} & \multicolumn{1}{c|}{$r_Q = 1/2$} & \multicolumn{1}{c|}{$r_Q = 2/3$} & \multicolumn{1}{c|}{$r_Q = 3/4$}\\
\hline
\hline
$k = 500$ & 0.051 & 0.024 & 0.014 & 0.032 & 0.014 & 0.011\\
\hline
$k = 1000$ & 0.055 & 0.028 & 0.018 & 0.039 & 0.020 & 0.014\\
\hline
$k = 2000$ & \textbf{0.058} & \textbf{0.030} & \textbf{0.021} & \textbf{0.045} & \textbf{0.023} & \textbf{0.016}\\
\hline
\end{tabular}
\end{table*}

\subsection{Goodput}
\label{Goodput}

When the quality of the quantum channel starts degrading, a QSC having a certain quantum coding rate $r_Q$, which was previously capable of correcting all the quantum errors flawlessly, may be no longer succeed in error-free decoding. In a condition where a QSC operates in the face of quantum channel having the depolarizing probability beyond its error correction capability, the QSC may in fact inflict more quantum errors by correcting them in the wrong positions. Hence, the quantum coding rate of a QSC should be adjusted according to the quality of the quantum channel. The most intuitive way of improving the error correction capability of a QSC is to reduce its quantum coding rate, which means imposing more redundancy. Similarly, when the quality of the quantum channel starts improving, one can increase the quantum coding rate accordingly in order to reduce the overhead imposed by the QSC and hence, improve the effective throughput.

To elaborate a little further, for a random QSC $\mathcal{C}$ exhibiting quantum coding rate of $r_Q$ having a sufficiently high number of physical qubits, there exists a limit $p^{\ast}$ below which it can operate perfectly yielding an infinitesimally low QBER. Hence, the goal of designing a QSC is to ensure that it can operate as close as possible to the limit of $p^{\ast}$. Similarly, for a given depolarizing probability $p$, we can find a random QSC $\mathcal{C}$ exhibiting a quantum coding rate of $r_Q \leq C_Q(p)$ and having a sufficiently high number of physical qubits that is capable of yielding an infinitesimally low QBER. This specific limit is referred to as the quantum hashing bound, which is defined as follows~\cite{bennett1996mixed, mackay2004sparse, wilde2014entanglement}:
\begin{equation}
\mathcal{C}_Q = 1 - H(p) - p \cdot \log_2(3),
\label{eq:hashing}
\end{equation}
where $H(p)$ is the binary entropy of $p$ defined by $H(p) = -p \log_2p - (1-p)\log_2(1-p)$. Given a value of $p$, then $C_Q$ is the quantum hashing bound for $p$. Conversely, given a value of $r_Q$, the value of $p^{\ast} = p(r_Q)$ represents the quantum hashing bound for $r_Q$. For instance, a QSC having a quantum coding rate of $r_Q = 1/2$, the quantum hashing bound is given by $p^{\ast} = 0.074$. In Fig.~\ref{fig:compare1}, the quantum hashing bound is represented by red dashed line. It can be observed that for $\text{QBER} = 10^{-3}$, the QSBC-QURC scheme operates closer to the quantum hashing bound. Quantitatively, the distance from the quantum hashing bound can be formally defined as
\begin{equation}
D \triangleq p^{\ast} - p,
\end{equation}
where $p$ is the achievable depolarizing probability, below which the QSC yields an infinitesimally low QBER. For example, in this treatise, we set the $\text{QBER} = 10^{-3}$. Therefore, based on the results in Table~\ref{table:results1}, the half-rate QIrCC-QURC scheme having $n = 2000$ physical qubits operates at $D = 0.074 - 0.037 = 0.0037$ from the quantum hashing bound, while the half-rate QSBC-QURC scheme also having $n = 2000$ physical qubits operates at $D = 0.074-0.045 = 0.029$ from the quantum hashing bound, provided that we have $\text{QBER} = 10^{-3}$. The distance from the quantum hashing bound constitutes a fair metric of comparing the efficiency of QSCs exhibiting various quantum coding rates. Hence, it has a direct relationship with the goodput, which represents the effective number of logical qubits after the decoding step. The achievable goodput taking quantum coding rate $r_Q$ into account for normalization can be formally defined as
\begin{equation}
\text{Goodput} = r_Q \cdot (1 - \text{QBER}).
\label{eq:goodput}
\end{equation}

By applying Eq.~\eqref{eq:goodput}, we can transform the QBER perfomance seen in Fig.~\ref{fig:compare3} into the goodput performance of Fig.~\ref{fig:compare5}. We have also plotted the quantum hashing bound formula of Eq.~\eqref{eq:hashing} to show the visual representation of the relationship between the quantum hashing bound and the goodput performance. In Fig.~\ref{fig:compare5}, the distance from the quantum hashing bound is shown by the red dashed lines for various quantum coding rates $r_Q$ at $\text{QBER} = 10^{-3}$. Quantitatively, given that $r_Q = \lbrace 1/2, 2/3, 3/4 \rbrace$, the resultant quantum hashing bound is given by $p^{\ast} = \lbrace 0.074, 0.044, 0.031 \rbrace$, while the distance from the quantum hashing bound is given by $D = \lbrace 0.029, 0.021, 0.015 \rbrace$.

\subsection{Reconfigurable Scheme}
\label{Reconfigurable Scheme}

Given a range of various requirements and quantum coding rates, the maximum tolerable depolarizing probability value, below which the QSBC-QURC schemes improve the QBER performance is portrayed in Table~\ref{table:results2}. As we have described earlier, for a certain requirement, there is a quantum coding scheme $\mathcal{C}$ that will satisfy it with the highest quantum coding rate. Again, for instance, given that the depolarizing probability of the quantum channel is $p = 0.01$ and the QBER requirement of $\text{QBER} < 10^{-3}$ is sufficient for the quantum computation or communication considered, we do not necessarily invoke a half-rate QSBC-QURC scheme for this purpose, since a $3/4$-rate QSBC-QURC scheme is already capable of satisfying the aforementioned conditions. By utilizing a $3/4$-rate QSBC-QURC scheme, we can have $50\%$ less auxiliary qubits. For QSBC-QURC scheme, a multi-rate scheme can be readily constructed since a single QSBC-QURC encoder is capable of providing multiple quantum coding rates, as described in Section~\ref{Quantum Short-Block Codes}.

\begin{figure}[ht!]
\center
\includegraphics[width=\linewidth]{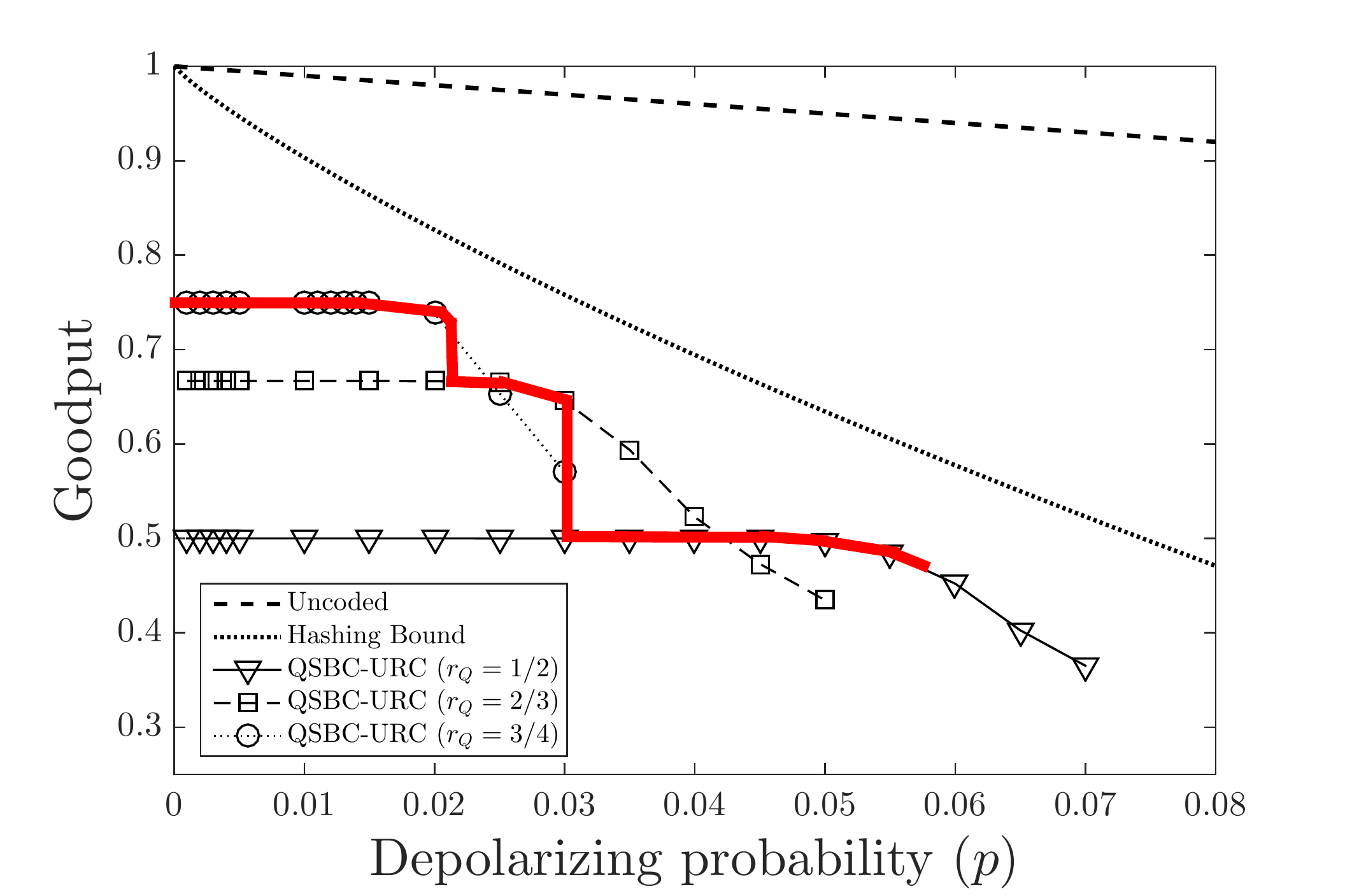}
\caption{Goodput versus depolarizing probability curves for QSBC-QURC schemes exhibiting various quantum coding rates having $k = 2000$ logical qubits after 16 decoding iterations. The bold red line represents the achievable goodput using our multi-rate scheme given the minimum requirement $ \text{QBER} \leq \text{QBER}_{\text{uncoded}}$.}
\label{fig:goodput1}
\end{figure}

Consequently, based on the goodput results of Fig.~\ref{fig:compare5}, we can determine the quantum coding rate switching point for our QSBC-QURC scheme in order to adjust the quantum coding rate based on the depolarizing probability experienced. For instance, the switching depolarizing probability for the requirement of $\text{QBER} < \text{QBER}_{\text{uncoded}}$ is given by $p = \lbrace 0.058, 0.030, 0.021 \rbrace$ for the three QSBC-QURC schemes considered, which can be seen in Table~\ref{table:results2}. The effective goodput after applying the switching probability is represented by the bold red line in Fig.~\ref{fig:goodput1}. We can also infer from Fig.~\ref{fig:goodput1} that for a depolarizing probability of $0.030 \leq p \leq 0.058$ we can utilize the QSBC-QURC having $r_Q = 1/2$ in order to maintain the requirement of $\text{QBER} < \text{QBER}_{\text{uncoded}}$, while the QSBC-QURC having $r_Q = 2/3$ is invoked for $0.021 \leq p \leq 0.030$, and finally a QSBC-QURC having $r_Q = 3/4$ is invoked for $p < 0.021$. 

\begin{figure}[ht!]
\center
\includegraphics[width=\linewidth]{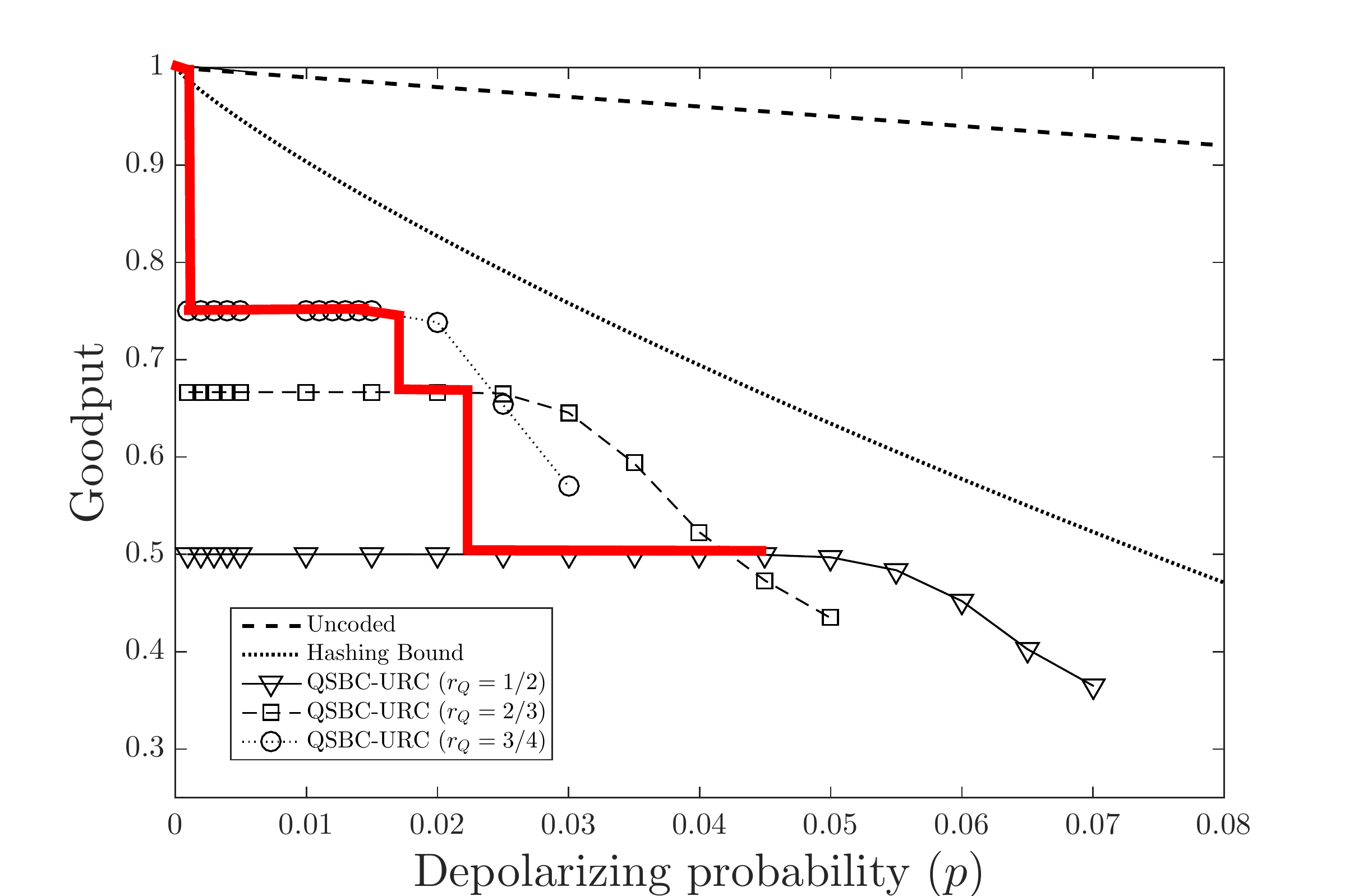}
\caption{Goodput versus depolarizing probability curves for QSBC-QURC schemes exhibiting various quantum coding rates having $k = 2000$ logical qubits after 16 decoding iterations. The red line represents the achievable goodput using our multi-rate scheme given the minimum requirement $ \text{QBER} = 10^{-3}$.}
\label{fig:goodput2}
\end{figure}

Similarly, the effective goodput attained upon applying the switching regime specified in Table~\ref{table:results2} for maintaining $\text{QBER} < 10^{-3}$ is portrayed in Fig.~\ref{fig:goodput2}. Based on Table~\ref{table:results2} and Fig.~\ref{fig:goodput2}, the QSBC-QURC schemes may be configured for operating at $r_Q = 1/2$ for $0.023 \leq p \leq 0.045$, operating at $r_Q = 2/3$ for $0.016 \leq p \leq 0.023$, operating at $r_Q = 2/3$ for $10^{-3} \leq p \leq 0.016$ and finally, the system may switch to the uncoded mode when the quantum channel reaches the condition of $p < 10^{-3}$.

\section{Conclusions and Future Research}
\label{Conclusions and Future Research}

We have conceived a QTC scheme exhibiting multiple quantum coding rates using a single quantum encoder. This construction was created by exploiting the inherent property of our QSBC encoders. We have amalgamated the QSBCs and QURC schemes in order to transform the error detection capability of QSBCs into an error correction capability without sacrificing the quantum coding rate. We predicted and analyzed the performance of our QSBC-QURC schemes both by EXIT chart analysis and by Monte Carlo simulations. Despite its low complexity, the QSBC-QURC schemes are capable of operating relatively close to the quantum hashing bound at $\text{QBER} = 10^{-3}$. Furthermore, we have compared our half-rate QSBC-QURC with the best performing half-rate QTC scheme, namely the QIrCC-QURC of~\cite{babar2016serially}. The QSBC-QURC outperforms the QIrCC-QURC in terms of operating closer to the quantum hashing bound at $\text{QBER} = 10^{-3}$, but observe in Fig.~\ref{fig:compare1} that the QIrCC-QURC has the edge over the QSBC-QURC for low values of $p$ due to the relatively high error floor of QSBC-QURC, which was indeed expected from its EXIT chart analysis.

We have also extended our discussions to the option of creating a multi-rate scheme for our QSBC-QURC. By using the distance from quantum hashing bound and the goodput, we quantified the normalized performance of our QSBC-QURC scheme by taking into account its quantum coding rate. Furthermore, we have also determined the quantum coding rate switching point based on the depolarizing probability for two specific requirements, $\text{QBER} < \text{QBER}_{\text{uncoded}}$ and $\text{QBER} < 10^{-3}$. Finally, we quantified the goodput achieved by adapting to the quantum depolarizing probability. 

As an initial study and first instantiation of QTCs exhibiting multiple quantum coding rates in a single quantum encoder, we present several potential research direction as an extention of this result. Firstly, it is indeed possible to have a higher quantum coding rate than $r_Q = 3/4$ as we have demonstrated in this treatise. However, the simulations will become more time consuming, since the simulation time of each block of the quantum encoder is roughly doubled each time we add one more qubit into the block. Our powerful parallel computer is only capable of simulating the QSBC-QURC exhibiting quantum coding rates up to $r_Q = 3/4$. Ultimately, the QSBCs can be combined to create a QSBC-QURC exhibiting an arbitrary quantum coding rate, similar to the QIrCCs. An EXIT-chart-based heuristic search can also be certainly conducted to yield the best combination of the subcomponent QSBCs. Consequently, this would result in a very smooth goodput performance curve.

However, one of the requirements for the multi-rate scheme for QSCs to work flawlessly is having perfect channel estimation for predicting the depolarizing probability. Our simulation results are based on the assumption that we have perfect knowledge of the depolarizing probability at the decoder. However, it is worth mentioning that it has been demonstrated in classical settings that classical parallel turbo codes are generally rather insensitive to innacurate signal-to-noise ratio (SNR) estimation~\cite{summers1998snr, worm2000turbo, khalighi2003effect}. However, for QTCs, the effect of inaccurate depolarizing probability knowledge on the QBER performance is still unknown. Some investigations towards the inaccuracy problem have been conducted for quantum low density parity check (QLDPC) codes in~\cite{xie2012channel, xie2016improved}. These results give an early indication that there is no significant QBER performance difference variation between having a perfect and imperfect quantum channel knowledge~\cite{fujiwara2014instantaneous}. Hence, the next step is to find the most appropriate depolarizing probability estimator for QSBCs. A plausible option is using some known pilot qubits or pre-shared entanglement for estimating the depolarizing probability. However, we have to rely on the idealistic assumption that the quantum depolarizing channel is a static channel implying that the value of depolarizing probability of the quantum channel does not vary much over time~\cite{bschorr2001channel, collins2015depolarizing, dumitrescu2015direct}. An alternative approach is invoking syndrome-based depolarizing probability estimation, which can be derived from the classical realm~\cite{toto2011maximum, lechner2013estimating, fujiwara2014instantaneous, combes2014situ}. The syndrome-based estimator may be deemed to be more efficient because it eliminates the need for pilot qubits or pre-shared entanglement and also the necessity of measuring the quantum information. However, a syndrome-based channel estimator would require a high number of physical qubits in order to work accurately whereas currently, we are working on a QSC scheme having a relatively short block $(k < 2000)$. Needless to say, a joint study of the effect of the inaccuracy depolarizing probability on the QBER performance of our QSBC-QURC scheme and the design of online depolarizing probability estimation is a promising subject. 

\bibliographystyle{ieeetr}

\clearpage

\begin{IEEEbiography}[{\includegraphics[width=1in,height=1.25in,clip,keepaspectratio]{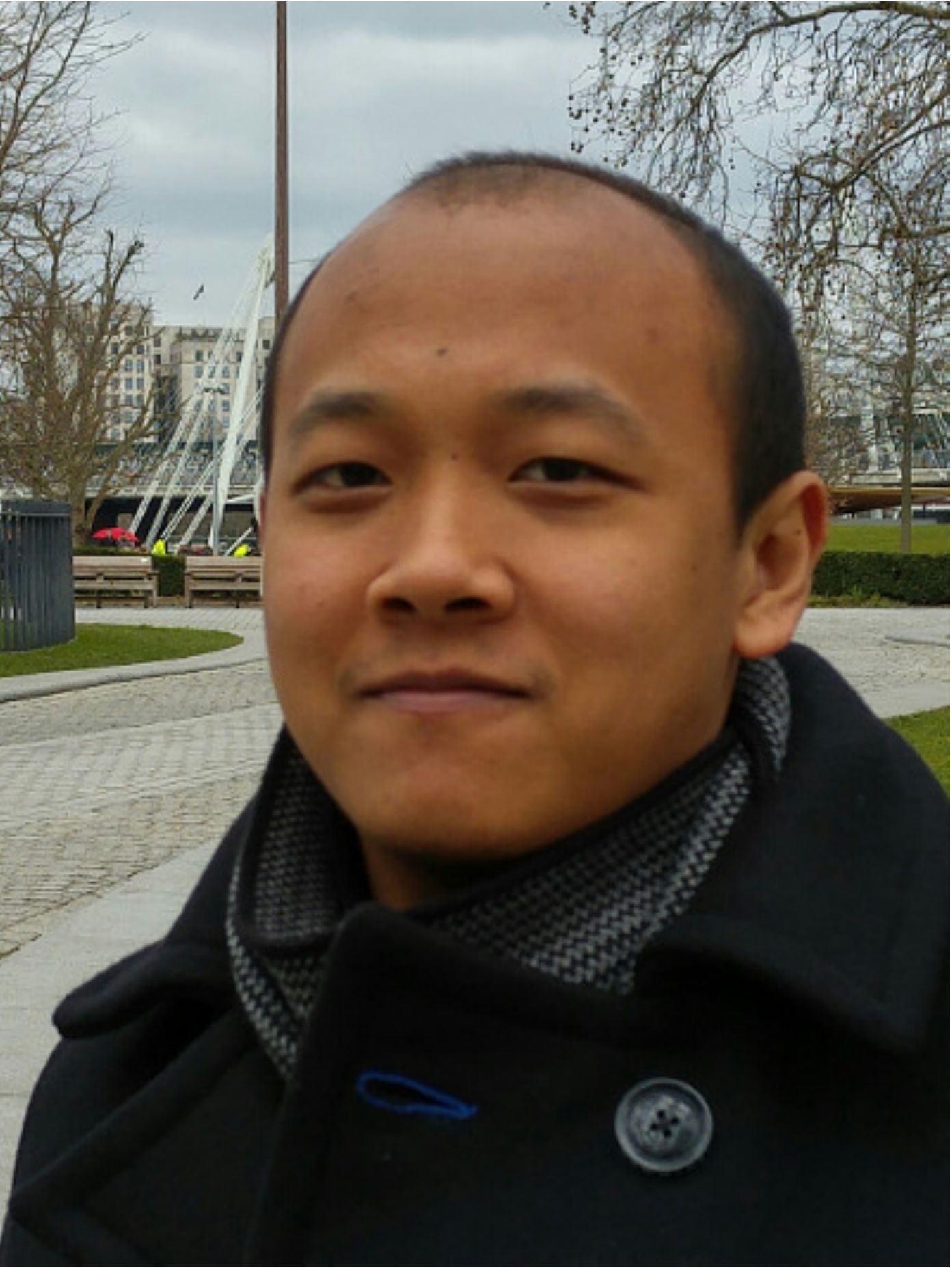}}]{Daryus Chandra} (S'15) received the M.Eng. degree in electrical engineering from Universitas Gadjah Mada, Indonesia, in 2014. He is currently pursuing the Ph.D. degree with the Next Generation Wireless Research Group, School of Electronics and Computer Science, University of Southampton, UK.

His research interests include classical and quantum error correction codes, quantum information, and quantum communications. He is a recipient of scholarship award from the Indonesia Endowment Fund for Education (Lembaga Pengelola Dana Pendidikan, LPDP). 
\end{IEEEbiography}

\begin{IEEEbiography}[{
\includegraphics[width=1in,height=1.25in,clip,keepaspectratio]{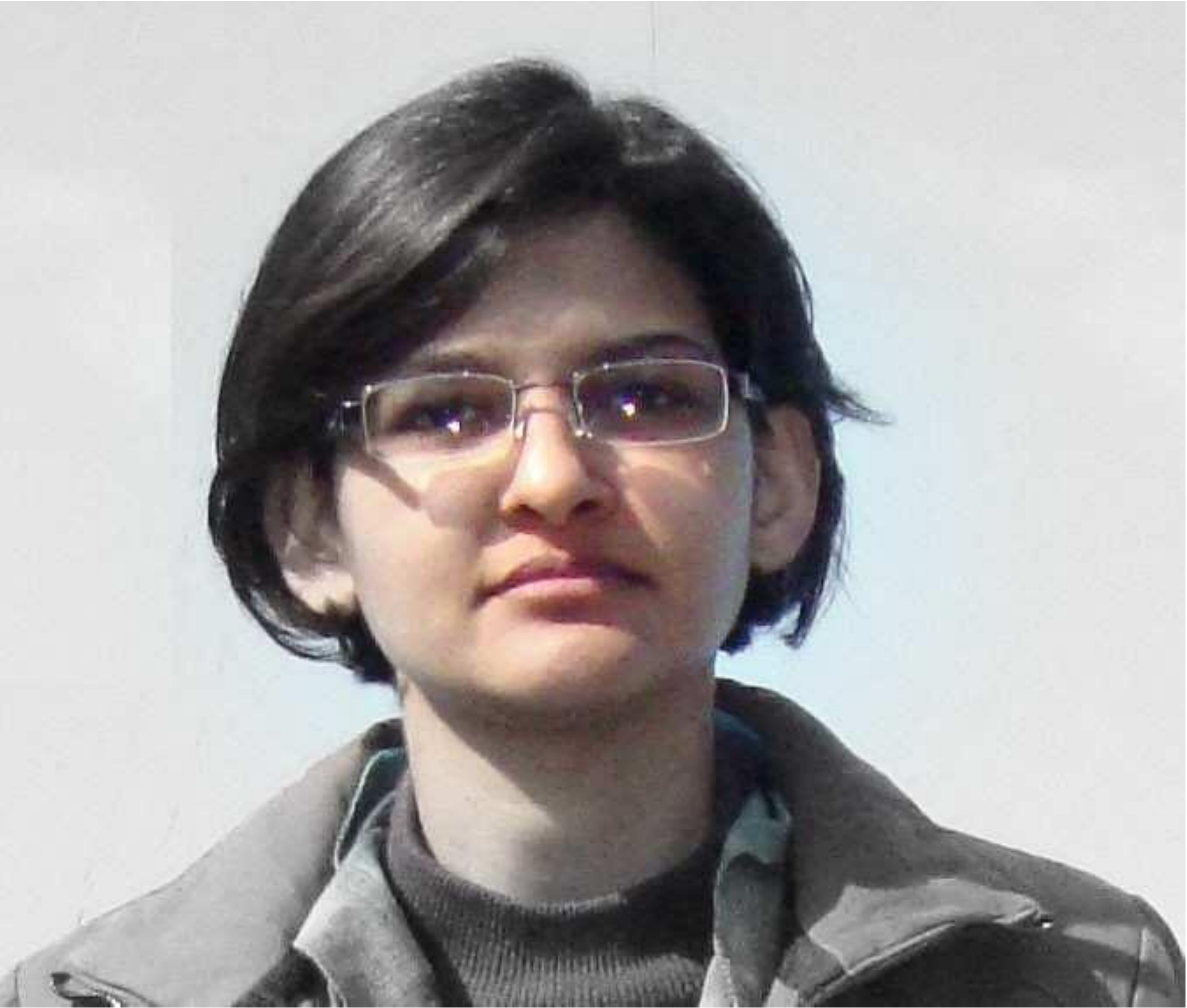}}]{Zunaira Babar} received the B.Eng. degree in electrical engineering from the National University of Science and Technology, Islamabad, Pakistan, in 2008, and the M.Sc. degree (Hons.) and the Ph.D. degree in wireless communications from the University of Southampton, U.K., in 2011 and 2015, respectively. She is currently a Research Fellow with the Next Generation Wireless Research Group, University  of Southampton. 

Her research interests include classical and quantum error correction codes, coded modulation, joint source and channel coding, error reconciliation for quantum key distribution, quantum-assisted communications, and optical communications. She was a recipient of several academic
awards, including the Commonwealth Scholarship from the Government of U.K. (2010-2011) and the Dean's Award for Early Career Research Excellence from the University of Southampton (2018).
\end{IEEEbiography}

\begin{IEEEbiography}[{
\includegraphics[width=1in,height=1.25in,clip,keepaspectratio]{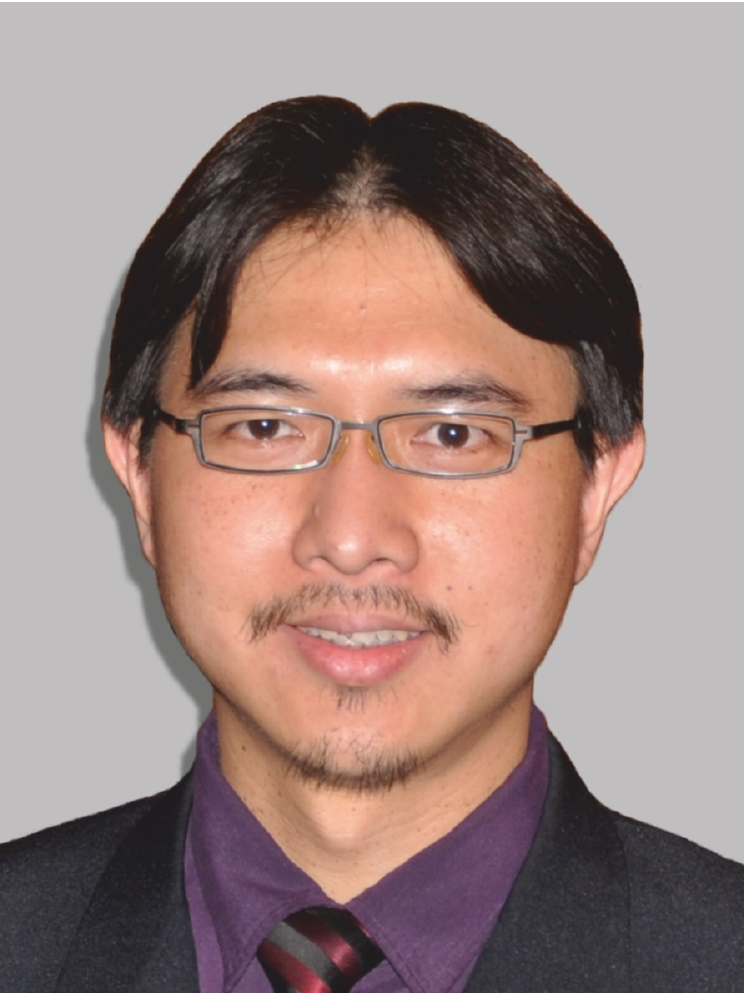}}]{Soon Xin Ng} (S'99-M'03-SM'08) received the B.Eng. degree (First Class) in electronic engineering and the Ph.D. degree in telecommunications from the University of Southampton, Southampton, U.K., in 1999 and 2002, respectively. From 2003 to 2006, he was a Post-Doctoral Research Fellow working on collaborative European research projects known as SCOUT, NEWCOM and PHOENIX. Since August 2006, he has been an Academic Staff Member with the School of Electronics and Computer Science, University of Southampton. He was involved in the OPTIMIX and CONCERTO European projects as well as the IU-ATC and UC4G projects. He was the Pricipal Investigator of an EPSRC project on Cooperative Classical and Quantum Communications Systems. He is currently an Associate Professor in telecommunications with the University of Southampton.

His research interests include adaptive coded modulation, coded modulation, channel coding, space-time coding, joint source and channel coding, iterative detection, OFDM, MIMO, cooperative communications, distributed coding, quantum communications, quantum error correction codes, and joint wireless-and-optical-fibre communications. He has published over 240 papers and co-authored two Wiley/IEEE Press books in the above areas. He is a Fellow of the Higher Education Academy in the UK, a Chartered Engineer, and a fellow of IET.
\end{IEEEbiography}

\begin{IEEEbiography}
[{\includegraphics[width=1in,height=1.25in,clip,keepaspectratio]{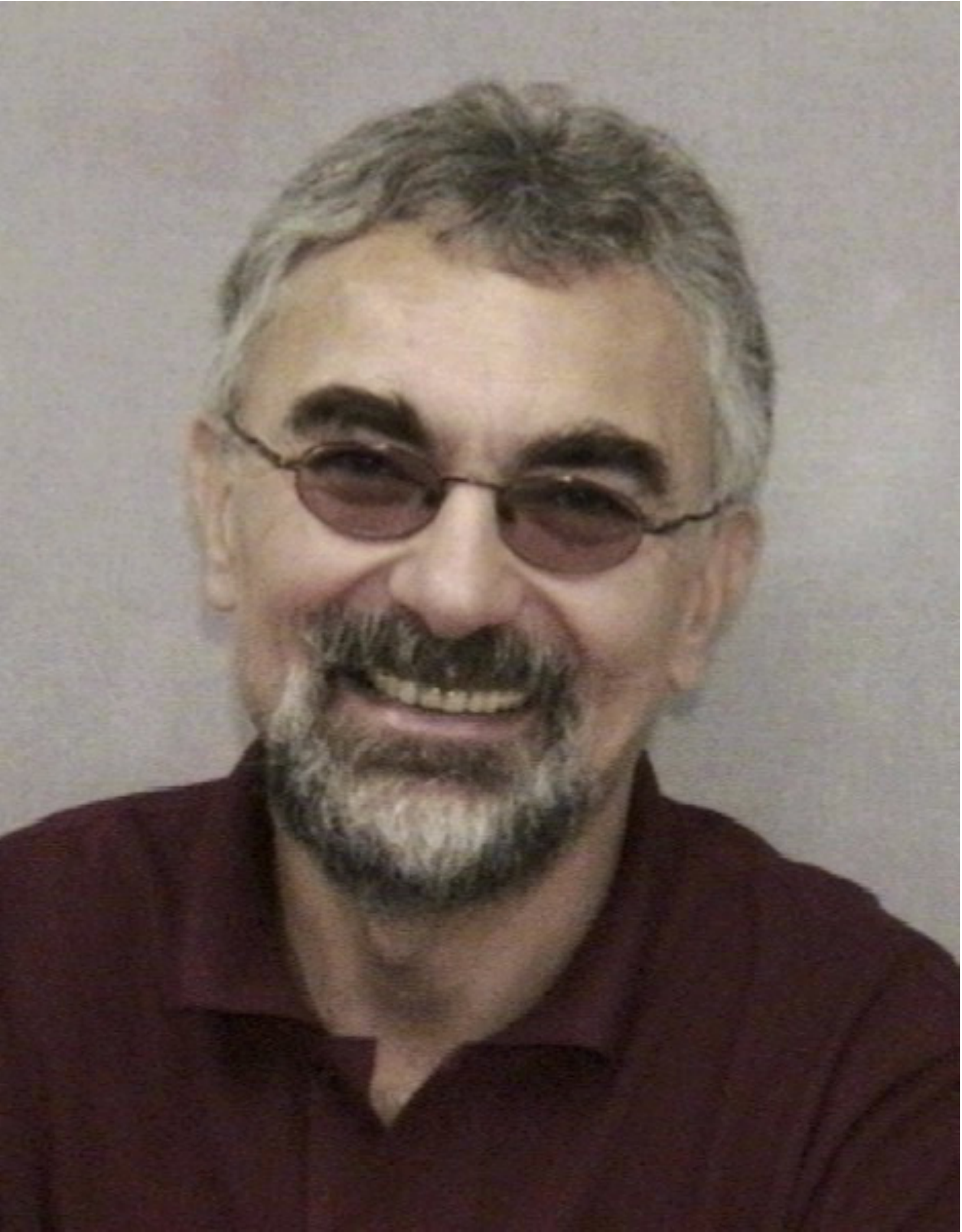}}] {\bf Lajos Hanzo} (M'91-SM'92-F'04) received his degree in electronics in 1976 and the Doctorate degree in 1983, and the Honorary Doctorate degrees (\textit{Doctor Honoris Causa}) from the Technical University of Budapest in 2009 and the University of Edinburgh in 2015. During his 40-year career in telecommunications he has held various research and academic posts in Hungary, Germany and the UK. Since 1986, he has been with the School of Electronics and Computer Science, University of Southampton, UK, where he holds the chair in telecommunications. He has successfully supervised 112 PhD students, co-authored 18 Wiley/IEEE Press books on mobile radio communications totalling in excess of 10~000 pages, published 1768 research contributions at IEEE Xplore, acted both as TPC and General Chair of IEEE conferences, presented keynote lectures and has been awarded a number of distinctions. He is currently directing a 40-strong academic research team, working on a range of research projects in the field of wireless multimedia communications sponsored by industry, the Engineering and Physical Sciences Research Council, U.K., the European Research Council’s Advanced Fellow Grant and the Royal Society’s Wolfson Research Merit Award. He is an enthusiastic supporter of industrial and academic liaison and he offers a range of industrial courses.

He was the Editor-in-Chief of the IEEE Press and a Chaired Professor also with Tsinghua University, Beijing, from 2008-2012. He is also a Governor of the IEEE ComSoc and of IEEE VTS. He is a fellow of the Royal Academy of Engineering, the Institution of Engineering and Technology, and the European Association for Signal Processing. For further information on research in progress and associated publications please refer to \url{http://www.wireless.ecs.soton.ac.uk}.
\end{IEEEbiography}

\end{document}